\documentclass[fleqn,usenatbib]{mnras}

\usepackage{newtxtext,newtxmath}

\usepackage[T1]{fontenc}
\usepackage{ae,aecompl}

\usepackage{graphicx}	
\usepackage{amsmath}	
\usepackage{natbib}     
\usepackage{longtable}  
\usepackage{CJK}        

\newcommand{\kms}{km\,s$^{-1}$}

\newcommand{\mum}{$\mu$m}
\newcommand{\Msun}{M$_\odot$}
\newcommand{\Msunyr}{M$_\odot$\,yr$^{-1}$}

\newcommand{\um}{\,\hbox{$\mu$m}}

\newcommand{\nH}{\,\hbox{$n_{\mathrm{H}}$}}

\newcommand{\ha}{\hbox{H$\alpha$}\,}

\newcommand{\cii}{\hbox{[C II] 158}\,\,}

\newcommand{\vSix} {$v_{16}$\,}
\newcommand{\vFive} {$v_{50}$\,}
\newcommand{\vEight} {$v_{84}$\,}
\newcommand{\wOne}{$W_{1\sigma}$\,}

\newcommand{\vFiveDelt} {$\Delta v_{50}$\,}
\newcommand{\wOneDelt}{$\Delta W_{1\sigma}$\,}

\newcommand{\barolo}{\textsc{$^{\mathrm{3D}}$barolo}}

\title[Dust Content of Galactic Halos. IV. NGC~3079.]{Exploring the Dust Content of
  Galactic Halos with {\it Herschel}. \\ IV. NGC~3079}

\author[S. Veilleux et al.]{
S. Veilleux,$^{1,2}$ \thanks{E-mail: veilleux@astro.umd.edu}
M. Mel\'{e}ndez,$^{1,3}$
M. Stone,$^{1}$
G. Cecil,$^{4}$
E. Hodges-Kluck,$^{5}$
\newauthor
J. Bland-Hawthorn,$^{6}$
J. Bregman,$^{7}$
F. Heitsch,$^{4}$
C. L. Martin,$^{8}$
T. Mueller,$^{9}$
\newauthor
D. S. N. Rupke,$^{10}$
E. Sturm,$^{9}$
R. Tanner,$^{4,11}$
C. Engelbracht$^{12}$
\\
$^{1}$Department of Astronomy, University of Maryland, College Park, MD 20742, USA\\
$^{2}$Joint Space-Science Institute, University of Maryland, College Park, MD 20742, USA\\
$^{3}$Space Telescope Science Institute, Baltimore, MD 21218, USA\\
$^{4}$Department of Physics and Astronomy, University of North Carolina, Chapel Hill, NC 27599, USA\\
$^{5}$X-ray Astrophysics Laboratory, NASA Goddard Space Flight Center, Greenbelt, MD 20771, USA\\
$^{6}$Department of Physics, University of Sydney, Sydney, NSW 2006, Australia\\
$^{7}$Department of Astronomy, University of Michigan, Ann Arbor, MI 48109, USA\\
$^{8}$Department of Physics, University of California, Santa Barbara, CA 93106, USA\\
$^{9}$Max-Planck-Institute for Extraterrestrial Physics (MPE), 85748 Garching, Germany\\
$^{10}$Department of Physics, Rhodes College, Memphis, TN 38112, USA\\
$^{11}$Universities Space Research Association, 7178 Columbia Gateway Dr., Columbia, MD 21046, USA\\
$^{12}$Department of Astronomy, University of Arizona, Tucson, AZ 85721, USA\\
}

\date{Accepted XXX. Received YYY; in original form ZZZ}

\pubyear{2021}

\begin{document}
\label{firstpage}
\pagerange{\pageref{firstpage}--\pageref{lastpage}}
\maketitle

\begin{abstract}
We present the results from an analysis of deep {\em Herschel} far-infrared
observations of the edge-on
disk galaxy NGC~3079.
The PSF-cleaned PACS images at 100 and 160 $\mu$m display a
25~$\times$~25~kpc$^2$ X-shape structure centered on the nucleus that
is similar in extent and orientation
to that seen in H$\alpha$, X-rays, and the far-ultraviolet. One of the
dusty filaments making up this structure is detected in the SPIRE
250~$\mu$m map out to $\sim$~25 kpc from the nucleus.
The match between the far-infrared filaments and those detected at
other wavelengths suggests that the dusty material has been lifted out
of the disk by the same large-scale galactic wind that has produced
the other structures in this object. A closer look at the central
10~$\times$~10~kpc$^2$ region provides additional support for this
scenario. The dust temperatures traced by the 100-to-160 $\mu$m flux
ratios in this region are enhanced within a biconical region centered
on the active galactic nucleus, aligned along the minor axis of the
galaxy, and coincident with the well-known double-lobed cm-wave radio
structure and H$\alpha$-X-ray nuclear superbubbles.
PACS imaging spectroscopy of the inner 6-kpc region reveals broad
[C~II] 158~$\mu$m emission line profiles and OH 79~$\mu$m absorption
features along the minor axis of the galaxy with widths
well in excess of those expected from beam smearing of the disk
rotational motion. This provides compelling evidence that the cool
material traced by the [C~II] and OH features directly interacts with
the nuclear ionized and relativistic outflows traced by the H$\alpha$,
X-ray, and radio emission.
\end{abstract}

\begin{keywords}
galaxies: halos -- galaxies: ISM -- galaxies: photometry -- galaxies:
starburst -- galaxies: star formation -- galaxies: infrared: galaxies
\end{keywords}



\section{Introduction} 
\label{sec:introduction}

There is growing evidence that galaxies are dynamically and chemically
evolving systems where gas flows in and out, or is processed into
stars \citep{Tumlinson2017,Peroux2020,Veilleux2020}.  Galaxies acquire
gas from galaxy mergers, galactic fountains, and intergalactic
accretion, but may also lose material in large-scale outflows driven
by stellar and black-hole driven processes. The dust, tied to the
outflowing gas by strong electrostatic forces, likely also
participates in this large-scale motion.  The amount of dust outside
of galaxies, inferred from reddening measurements of background
quasars and galaxies by foreground galaxy halos, is comparable to that
within galaxies \citep{Menard2010,Peek2015}, but the origin of this
dust is uncertain. Presumably made inside galaxies, this dust was
likely transported into the halos via tidal and ram-pressure stripping
or large-scale outflows. Direct evidence for dusty outflows that
extend on the scales of the circumgalactic medium (CGM; $\la$ 100 kpc)
and intergalactic medium (IGM; $\ga$ 100 kpc) 
remains elusive, although this is changing
\cite[e.g.,][]{Rupke2019,Hodges-Kluck2020,Burchett2021}.

The present paper is the fourth in a series that reports the results
from a program conducted with the {\em Herschel} Space Observatory
which examines the dust content of nearby galaxies with known galactic
winds. Earlier papers have targeted the edge-on star-forming disk
galaxies NGC~4631 \citep[][Paper I]{Melendez2015} and NGC~891
\citep[][Paper III]{Yoon2021}, and a number of star-forming dwarf
galaxies highlighted by NGC~1569 \citep[][Paper
  II]{McCormick2018}. The target of the present paper is NGC~3079, a
nearby ($z$ = 0.0037) edge-on \citep[$i = 82^\circ$;][]{Veilleux1999}
disk galaxy with multi-wavelength evidence for a bisymmetric nuclear
outflow. Warm ionized gas superbubbles
\citep{Ford1986,Filippenko1992,Veilleux1994,Cecil2001} with tightly
correlated hot ionized gas \citep{Cecil2002} have been observed along
the minor axis (P.A.\ = 75$^\circ$) of the galaxy disk, accompanied
with a figure-8/double-lobe relativistic radio structure
\citep{Duric1983,Irwin1988,Irwin2019,Sebastian2019} and extended
nonthermal hard X-ray emission \citep{Li2019}.

There is some ambiguity about the driving mechanism (starburst or AGN)
of this outflow.  There is plenty of molecular gas in the central core
of this galaxy to fuel star formation
\citep{Young1988,Sofue2001,Koda2002} and the implied central star
formation rate (SFR) from {\em Akari} \citep[2.6
  \Msunyr;][]{Yamagishi2010} may be sufficient to drive the outflow,
but this value is derived while ignoring the AGN contribution to the
infrared emission.  Excess 20 $-$ 100 keV continuum emission
\citep{Iyomoto2001} and strong 6.4 keV Fe K line emission
\citep{Cecil2002} point to the presence of an obscured AGN with a high
hydrogen column density \citep[$N_{\rm H} \approx 10^{24.4 - 24.6}$
  cm$^{-2}$;][]{Brightman2015,Masini2016,Ricci2017}, consistent with
the picture of a nearly edge-on nuclear molecular disk responsible for
the H$_2$O megamaser on parsec scale \citep{Kondratko2005}.  A radio
continuum emitting jet at centimeter wavelengths is seen working its
way out of the innermost 1 to 2 pc clumpy region around the central
AGN
\citep{Irwin1988,Trotter1998,Yamauchi2004,Middelberg2007}. \citet{Cecil2001}
have shown that the very broad optical line emission at the base of
the NE H$\alpha$ bubble is positioned along the position angle of the
inner radio jet, and thus have favored a scenario where the kinetic
energy from the AGN jet is deposited (thermalized) at the base of the
bubble and ultimately drives the observed biconical outflow \citep[see
  also][]{Middelberg2007}. This same outflow, regardless of its exact
origin, is almost certainly responsible for the broad but shallow
blueshifted (by up to 600 \kms) H~I 21-cm absorption in this system
\citep{Shafi2015}.


The bisymmetric outflow of NGC~3079 has long been suspected to extend
beyond 10 kpc based on the detections of faint large-scale extraplanar
optical-line filaments \citep{Heckman1990,Veilleux1995,Cecil2001},
long-wavelength radio emission \citep{Irwin2003}, and soft X-ray
structures \citep{Fabbiano1992,Strickland2004a}. The cometary H~I 21-cm
plume of the companion NGC~3073, located within the western biconical
outflow $\sim$ 50 kpc from NGC~3079 \citep{Irwin1987,Shafi2015}, has
provided indirect evidence that the outflow may extend considerable
further. This was recently confirmed by \citet{Hodges-Kluck2020}, who
examined deep archival {\em XMM-Newton} X-ray images and {\em GALEX}
NUV and FUV maps and revealed the existence of a 60-kpc biconical
structure of hot ionized and dusty material aligned along the central
kpc-scale outflow.


The main objective of the present paper is to examine the cool ($T
\simeq$ 10 $-$ 1000 K) dust and gas associated with both the small-
and large-scale outflows in NGC~3079. We first present deep {\em
  Herschel} far-infrared images that trace the continuum emission from
dust on large scale, and then search for cool gas impacted by the
nuclear outflow using far-infrared imaging spectroscopy of [C~II]
157.74 $\mu$m (\cii for short) and the OH 119.233, 119.441 $\mu$m
doublet (OH 119 for short). Section \ref{sec:observations} describes
the observations and data processing.  Section \ref{sec:results}
presents the main results. Section \ref{sec:discussion} provides a
more detailed and quantitative analysis of these results, and compares
the {\em Herschel} data with some of the latest data at other
wavelengths, on both small and large scales. The results of these
comparisons are discussed against theoretical predictions of dust and
cool gas entrainment and in the more general context of galaxy
evolution and galaxy ecosystems.  Section \ref{sec:conclusions}
summarizes the main conclusions. In this paper we adopt a distance of
19 Mpc for NGC~3079 \citep{Springob2007}, which corresponds to a scale
of $\sim$ 91 pc per arcsecond.


\section{Observations and Data Processing}
\label{sec:observations}

\subsection{Deep Far-Infrared Imaging}
\label{sec:FIR_imaging}



NGC~3079 was imaged with both the Photodetector Array Camera and
Spectrometer \citep[PACS;][]{Poglitsch2010} and the Spectral and
Photometric Imaging Receiver \citep[SPIRE;][]{Griffin2010} on board
{\em Herschel} as part of our cycle 1 open-time program
(OT1\_sveilleu\_2, PI: S.\ Veilleux; Table
\ref{tab:observations}). For the PACS observations, we obtained
simultaneous imaging in the PACS ``green'' 100 $\mu$m (85 $-$ 130
$\mu$m) and ``red'' 160 $\mu$m (130 $-$ 210 $\mu$m) channels in scan
mode along six position angles at 55$^\circ$, 70$^\circ$, 85$^\circ$,
95$^\circ$, 110$^\circ$, and 125$^\circ$ (Obs IDs: 1342231542,
1342231544, 1342231545, 1342231546, 1342231547, and 1342231543,
respectively). At each orientation angle, we requested 60 scan legs of
4\farcm0 length, 4\farcs0 scan leg separation, a repetition factor of
two, and a scan speed of 20\arcsec~s$^{-1}$. The total time per
position angle, including telescope overhead, was $\sim$ 1.6 hours,
for a total request of 9.3 hours. The SPIRE observations (Obs ID:
1342221916 from the same program) were taken simultaneously at 250,
350, and 500 $\mu$m in large map mode covering an area of
30\arcmin\ $\times$ 30\arcmin\ with two repetitions at nominal speed
(30\arcsec~s$^{-1}$).


\begin{table*}
	\centering
\caption{Summary of the Observations}
	\label{tab:observations}
	\begin{tabular}{ccccc} %
		\hline
Instrument/mode & Waveband [\mum] &
t$_{\rm exp}$[s] & Obs ID & Program ID\\
		\hline
(1) & (2) & (3) & (4) & (5)\\
		\hline
PACS/photometer & 100 & 33576 & 1342231542-47 & OT1\_sveilleu\_2\\
PACS/photometer & 160 & 33576 & 1342231542-47 & OT1\_sveilleu\_2\\
SPIRE/photometer & 250 & 2929 & 1342221916 & OT1\_sveilleu\_2\\
SPIRE/photometer & 350 & 2929 & 1342221916 & OT1\_sveilleu\_2\\
SPIRE/photometer & 500 & 2929 & 1342221916 & OT1\_sveilleu\_2\\
PACS/spectrometer & [C~II] 158, OH 119 & 8045 & 1342221391 & DDT\_esturm\_4\\
		\hline
	\end{tabular}
        \\
{Column (1): Instrument / mode of observation; Column
  (2): Waveband or central wavelength of the spectral scan in the
  rest-frame of NGC~3079 in $\mu$m; Column (3): Exposure time in
  seconds; Column (4): Observation ID; Column (5): Program ID. }
\end{table*}


The reduction of the PACS photometric data was done using the {\em
  Herschel} Interactive Processing Environment \citep[HIPE;][]{Ott2010}
version 8.0, following the exact same procedure as that used in Paper
I. We only summarize the important steps here and refer the reader to
Paper I for more details.  We followed the standard pipeline procedure
to convert from Level 0 to Level 1 data, including the extraction of
the calibration tree needed for the data processing, correction for
electronic crosstalk, application of the flat-field correction, and
finally deglitching and conversion from Volts to Janskys per array
pixel. To correct for bolometer drift (low frequency noise), both
thermal and non-thermal (uncorrelated noise), and to create the final
maps from the Level 1 data, we used the algorithm implemented in {\em
  Scanamorphos} \citep[v21.0;][]{Roussel2013}, which makes use of the
redundancy built in the observations to derive the brightness drifts.
The final PACS maps have a pixel size of $\sim$ 1/4 of the PSF
FWHM,
i.e.,
1\farcs7 at 100 $\mu$m and 2\farcs85 at 160 $\mu$m. {\em Scanamorphos}
also produces error and weight maps. The error map is defined as the
error on the mean brightness in each pixel.
Note that it does not include any error propagation associated with
the different steps performed on the pipeline. 
Given the relatively small field of view of our observations, relative
to the size of the galaxy, and the observing strategy, we created the
{\em Scanamorphos} maps with the ``minimap'' and ``flat'' options. On
the other hand, the SPIRE data was processed from level 0 up to level
1 with the HIPE scripts included in the {\em Scanamorphos}
distribution. The preprocessing by the pipeline includes the same
steps as in the PACS pipeline except that the conversion to brightness
is in Janskys per beam and the thermal drifts are subtracted by using
the smoothed series of thermistors located on the detector array as
the input of the drift model \citep[see][for details]{Ott2010}. To
convert the SPIRE data from Jy per beam to Jy per pixel, we used the
pipeline beam area as provided in the SPIRE Data Reduction Guide,
Table 6.9 (SPIRE-RAL-DOC 003248, 23 Nov 2016), i.e., 469.3542,
831.275, 1804.3058 arcsec$^2$ at 250, 350, and 500 $\mu$m,
respectively. The final SPIRE maps have a pixel size of $\sim$ 1/4 of
the PSF FWHM, i.e., 4\farcs5, 6\farcs25, and 9\farcs0 at 250, 350, and
500 $\mu$m, respectively.

\subsection{Far-Infrared Imaging Spectroscopy}
\label{sec:FIR_imaging_spectroscopy}


The PACS spectroscopic observations of NGC~3079 were obtained as part
of Director's Discretionary Time program DDT\_esturm\_4 (PI:
E.\ Sturm, Obs ID 1342221391; Table \ref{tab:observations}). These
observations were done in pointed observing mode with the PACS range
scan spectroscopy Astronomical Observing Template (AOT). A medium
chopper throw was used and the total on-source exposure time was 8045
seconds, resulting in a slightly irregular 5 $\times$ 5 grid of
9\farcs4 spaxels, where each spectrum covers redshifted [C~II] 158
$\mu$m and the OH 119 $\mu$m doublet, among other features, with a
spectral resolution $R \simeq$ 240 and angular resolution of $\sim$
11\farcs5 (estimated from Fig.\ 8 in the PACS Spectroscopy Performance
and Calibration Guide Issue 3.0).



The PACS spectroscopic data were retrieved from the {\em Herschel}
Science Archive
(HSA)\footnote{http://www.archives.esac.esa.int/hsa/whsa/} via HIPE
v15.0.0. These data have been pipeline-processed at the {\em Herschel}
Science Centre with the Standard Product Generation (SPG) software
v14.2.0 up to the \emph{rebinnedCube} task. The standard reduction
steps include glitch masking, bad and noisy pixel masking, dark and
background subtraction, spectral flatfielding, and flux calibration to
Jy per spaxel.

The [C~II] data cubes were interpolated onto a regular spatial grid of
1\arcsec\ spaxels.  For each spaxel in the [C~II] data, a second-order
polynomial was fit to the continuum emission and subtracted from the
spectrum. Two Gaussian components were then fit to the
continuum-subtracted spectrum and line profile properties are
estimated using the sum of the two components.  The uncertainty on the
absolute flux calibration is of the order $\sim 6$\% and the
wavelength calibration uncertainty for the red channel is $\sim$ 20
\kms\
\citep{Poglitsch2010}.  Note however that for pointed observations
the (original) spaxel size of 9\farcs4 is not small enough to
spatially sample the beam well (the beam FWHM is $\sim$
10\arcsec). The data are therefore undersampled and the true
morphology of an object will be degraded.
Nevertheless, a comparison of the published [C~II] fluxes,
  velocities, and line widths measured from the PACS [C~II] data of
  M~82 obtained in raster mode where the sky was Nyquist sampled
  \citep{Contursi2013} against those derived by fitting two Gaussians
  to the interpolated data of the same object and same line obtained
  in the pointed mode provides confidence in the use of such
  interpolated data cubes from pointed observations \citep[typically,
    the fluxes agree to within 10-25\% while the velocity centroids
    and line widths agree to within 10\% of the instrumental velocity
    resolution;][]{Stone2020}. 


%

\begin{figure*}
\centering
\includegraphics[width=\textwidth]{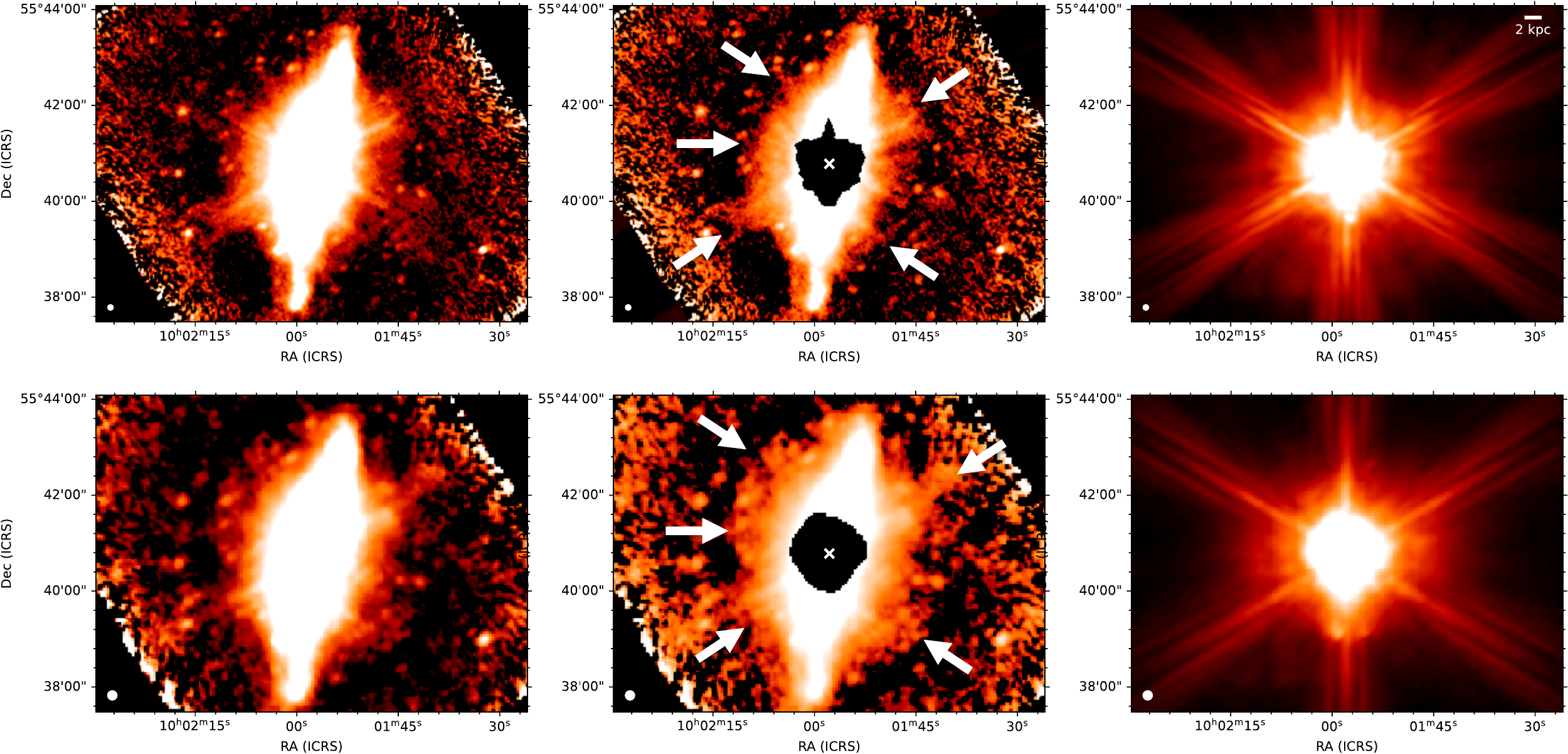} 
\caption{Deep PACS 100 $\mu$m (top row) and 160 $\mu$m (bottom row)
  images of NGC~3079. North is up and East to the left. The beam FWHM
  is displayed in the lower left corner of each panel. The linear
  scale is shown in the upper right panel. The left panels display the
  output from HIPE and {\em Scanamorphos}. The central panels show the
  results after removal of the diffraction spikes from the bright
  central source.
  The central region masked in black represents 95\% of the total flux
  of the reference PSF. The right panels show the final PSF
    produced by combining the instrumental PSF at each peak pixel
    found by the CLEAN algorithm (not on the same intensity scale as
    the left and middle panels). In other words, (left panel) =
    (middle panel) + (right panel). All panels are on an arcsinh
  intensity scale. The white arrows point to extraplanar features
  detected in the CLEANED images.}
\label{fig:ngc3079_pacs}
\end{figure*}

The OH data cube of NGC~3079 was reduced in the same manner as the
[C~II] data cube, except that it was not interpolated to smaller
spaxels to avoid erasing the abrupt changes in the absorption /
emission line profiles observed in the data cube from spaxel to spaxel
(discussed in Sec.\ \ref{sec:results_imaging_spectroscopy_oh}).
For each spaxel, a first-order spline was fit to the continuum
emission and subtracted from the spectra. Profile fitting of the OH
119 doublet followed a similar procedure from \citet{Veilleux2013} and
\citet{Stone2016}. The OH doublet profile is modeled using four
Gaussian components (two for each line of the doublet). The separation
between the two lines of the doublet was set to 0.208 \um\, in the
rest frame ($\sim$ 520 \kms) and the amplitude and standard deviation
were fixed to be the same for each component in the doublet.  The
uncertainties on the measurements of the fluxes, centroid velocities,
and velocity widths are estimated to be of the order $\sim$ 10\%, 25
\kms, and 50 \kms, respectively, in the strongest features, but are
considerably larger in the weaker features (see
Sec.\ \ref{sec:results_imaging_spectroscopy_oh}).



\section{Results}
\label{sec:results}

\subsection{PACS and SPIRE Far-Infrared Maps}
\label{sec:results_imaging}

The PACS 100 $\mu$m and 160 $\mu$m images of NGC~3079 are shown in the
left panels of Figure \ref{fig:ngc3079_pacs}. Diffraction spikes
associated with the central source are present in these data and need
to be removed before we can make any quantitative statement about the
extraplanar dust in this system.  To remove the diffraction spikes, we
follow the procedure detailed in \citet{McCormick2018}. We employ a
modified version of the CLEAN algorithm\footnote{Adapted from
  http://www.mrao.cam.ac.uk/$\sim$bn204/alma/python-clean.html.}
\citep{Hogbom1974}. Our CLEAN algorithm finds the peak pixel within an
area similar to the disk region, subtracts the appropriate PSF scaled
by a pre-defined gain as a fraction of the peak pixel value, and
repeats these two steps until the peak pixel value meets or drops
below a pre-defined minimum threshold value. Once the minimum
threshold value is reached, the algorithm outputs the component and
residual images. The reference PSFs for PACS\footnote{PACS Photometer
  Point Spread Function. Document PICC-ME-TN-033.} are chosen as our
beam PSFs.  The PSFs are rotated to match the galaxy observations,
centered on their central pixels, and scaled by normalizing their
central peak pixels to a value of unity for gain
multiplication. Finally, we apply our CLEAN algorithm to each of the
PACS and SPIRE images iteratively, lowering the threshold value to
determine circumgalactic flux value convergence.

Figure \ref{fig:ngc3079_pacs} illustrates the ``before'' (left
  panels) and ``after'' (middle panels) results of applying our
CLEAN algorithm to the PACS maps of NGC~3079. The diffraction spikes
(shown in the right panels) are more prominent at 100 $\mu$m
than at 160 $\mu$m, and become negligible in the SPIRE maps at 250
$\mu$m and beyond. This is a sign that the spectral energy
distribution (SED) of the central energy source (AGN + nuclear
starburst) in NGC~3079 is steeply declining at longer
wavelengths. This is different from our results on the star-forming
dwarf galaxies where no obvious trend was found with wavelength
\citep{McCormick2018}.

\begin{figure}
\centering
\includegraphics[width=0.35\textwidth]{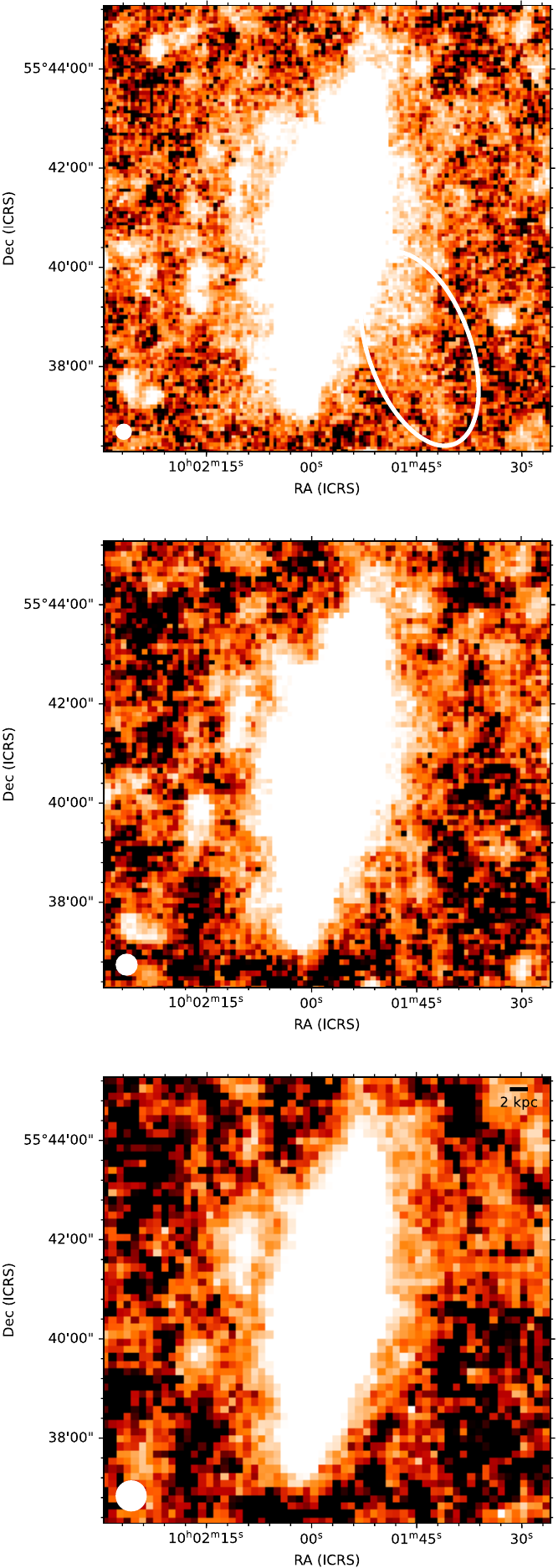} 
\caption{SPIRE 250 $\mu$m (top panel), 350 $\mu$m (middle), and 500
  $\mu$m (bottom) images of NGC~3079, after processing with {\em
    Scanamorphos}. North is up and East to the left. The beam FWHM is
  displayed in the lower left corner of each panel. The linear scale
  is shown in the top right corner of the bottom panel. All panels are
  on an arcsinh intensity scale. The white ellipse indicates the SW
  filament, securely detected only in the 250 $\mu$m image.}
\label{fig:ngc3079_spire}
\end{figure}

The cleaned 100 and 160 $\mu$m maps (middle panels in
  Fig.\ \ref{fig:ngc3079_pacs}) show a X-shape structure that is not
  associated with residuals from the PSF. This structure extends over
  a scale of $\sim$ 25 $\times$ 25 kpc$^2$ centered on the nucleus but
  is not pointing back to the nucleus. Instead the filaments originate
  at the base of the stellar disk at a galactocentric radius of $\sim$
  5 kpc on both sides from the nucleus. The filament in the SW
  quadrant is also detected in the SPIRE 250 $\mu$m image
  (Fig.\ \ref{fig:ngc3079_spire}), extending out $\sim$ 25 kpc from
  the nucleus along PA = 215$^\circ$ or $\sim$ 15 kpc from the
  mid-plane of the galaxy. The filament in the NW quadrant may also be
  detected in this image but it is within the confusion noise of the
  data.

\begin{figure}
\centering
\includegraphics[width=0.4\textwidth]{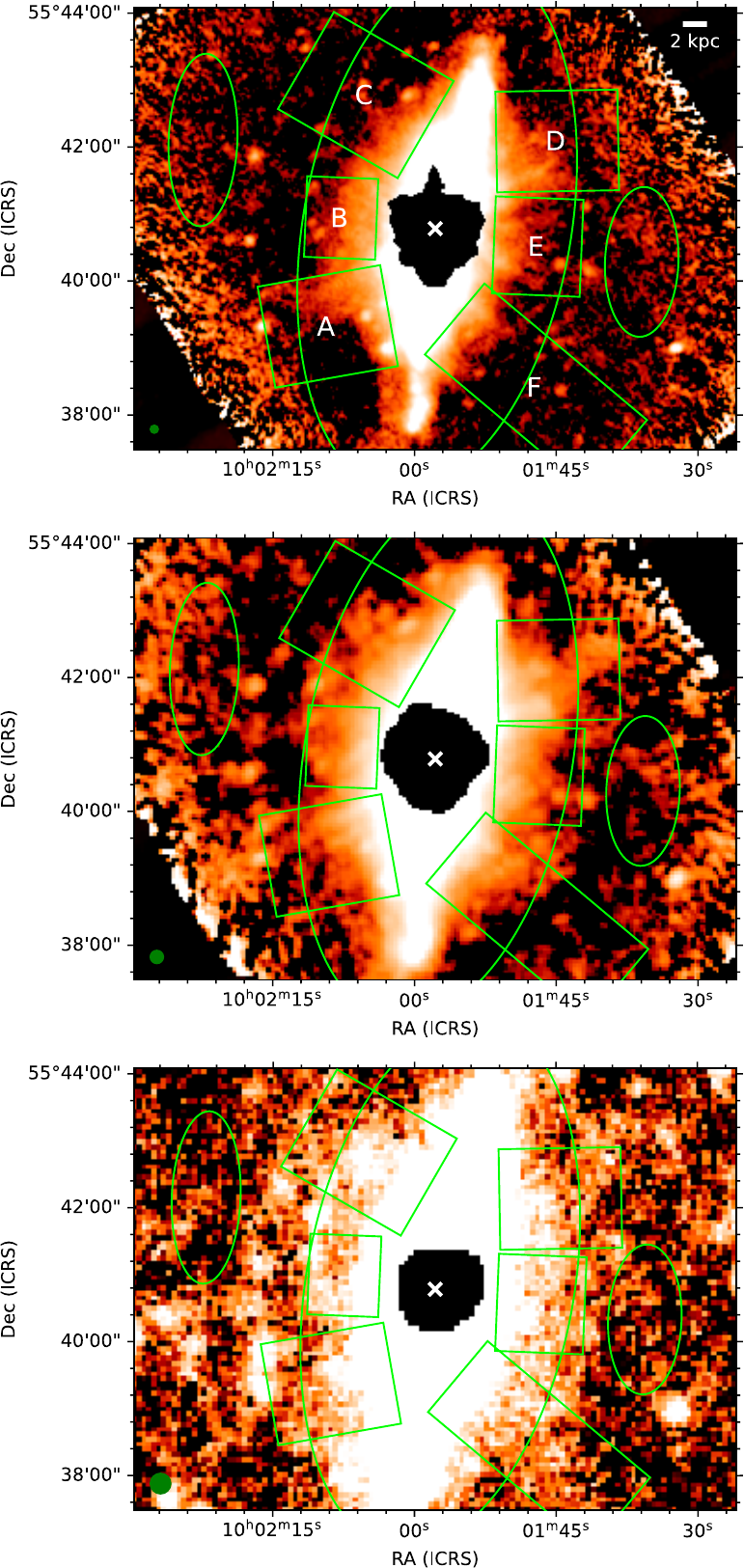}
\caption{Extraction apertures (green boxes) used to calculate the flux
  densities and corresponding dust masses and temperatures in the
  extraplanar dust features detected in the PACS 100 and 160 $\mu$m
  images (top and middle panels) and SPIRE 250 $\mu$m image (bottom
  panel). The large ellipse is used to calculate the global values for
  the entire galaxy, while the small ellipses indicate the regions
  where the sky background was calculated. North is up and East to the
  left. The beam FWHM is displayed in green in the lower left
  corner of each panel. The linear scale is shown in the top
  panel. The results of this analysis are listed in Table
  \ref{tab:dust_mass}. All panels are on an arcsinh intensity scale. }
\label{fig:ngc3079_apertures}
\end{figure}

Next, we estimate the far-infared fluxes emitted by these extraplanar
features. The results are listed in Table \ref{tab:dust_mass}.  While
the dust masses derived from these fluxes are discussed later in
Section \ref{sec:discussion_large_scale}, here we discuss how we
derived the fluxes. The selected apertures are sketched in Figure
\ref{fig:ngc3079_apertures}. Apertures A-F trace extraplanar dust
features while the large ellipse is used to derive global quantities.
The sky background (and standard deviation) was measured from the two
small elliptical apertures placed above and below the galaxy plane,
not too close to the galaxy (to avoid galaxy contamination) and far
from the edges of the map (to avoid unreliable fluxes due to elevated
noise). The total uncertainty on the integrated photometric
measurements is a combination of the error on the mean brightness in
each pixel added in quadrature within the source aperture (the error
map produced by {\em Scanamorphos}), the standard deviation of all the
pixels in the background aperture, and the PACS photometer flux
calibration accuracy. For the calibration uncertainties (extended
sources), we adopted a large (conservative) value of 10\% for both
PACS and SPIRE. This value comes from adding the systematic (4-5\%),
statistical (1-2\%), and PSF/beam size uncertainties (4\%). To derive
aperture corrections, we used the higher resolution image from
Spitzer/IRAC 8.0 $\mu$m \citep{McCormick2013} and measured the total
flux with the same aperture employed for our analysis. Then we
convolved the same image with the appropriate kernel to bring it to
the PACS
resolution\footnote{http://www.astro.princeton.edu/$\sim$ganiano/Kernels.html}
\citep{Aniano2011} and remeasured the flux in the same aperture. The
ratio of the unconvolved to the convolved (the same PSF as the {\em
  Herschel} PACS) flux is used as an estimate of the aperture
correction. For the global flux, the big aperture resulted in small
corrections with values of 1.01 and 1.02 for PACS 100, and 160 $\mu$m,
respectively, and 1.00 for all of the SPIRE images.
We color-corrected PACS fluxes assuming a modified blackbody with
$\beta$ = 2 and a blackbody temperature of $T$ = 20 K (0.974/0.971 at
100/160 $\mu$m).\footnote{PACS Photometer Passbands and Color
  Correction Factors for Various Source spectral energy distributions
  (SEDs). Document PICC-ME-TN-038, footnote \#4.}

\begin{table*}
	\centering
\caption{Dust Masses and Temperatures in the X-Shape Structure}
	\label{tab:dust_mass}
	\begin{tabular}{cccccccc} %
		\hline
Region & $f_\nu$(100 $\mu$m) & $f_\nu$(160 $\mu$m) & $f_\nu$(250 $\mu$m) & $f_\nu$(350 $\mu$m) & $f_\nu$(500 $\mu$m) & $T_D$ & log~$M_d$\\
& (Jy) & (Jy) & Jy) & (Jy) & (Jy) & (K) & (\Msun)\\
		\hline
(1) & (2) & (3) & (4) & (5) & (6) & (7) & (8)\\
		\hline
Global & 109.209 & 119.264 & 47.259 & 19.756 & 7.303 & 23.5 $\pm$ 0.63 & 8.0 $\pm$ 0.04\\
A & 0.302 & 0.737 & 0.524 & --- & --- & 16.9 $\pm$ 0.10 & 6.45 $\pm$ 0.02\\
B & 0.205 & 0.380 & 0.256 & --- & --- & 18.2 $\pm$ 0.03 & 6.02 $\pm$ 0.01\\
C & 0.198 & 0.543 & 0.535 & --- & --- & 15.6 $\pm$ 0.26 & 6.56 $\pm$ 0.04\\
D & 0.270 & 0.672 & 0.427 & --- & --- & 17.2 $\pm$ 0.26 & 6.35 $\pm$ 0.04\\
E & 0.166 & 0.442 & 0.366 & --- & --- & 16.2 $\pm$ 0.04 & 6.35 $\pm$ 0.01\\
F & 0.244 & 0.638 & 0.672 & --- & --- & 15.6 $\pm$ 0.39 & 6.65 $\pm$ 0.08\\
A $-$ F & 1.384 & 3.412 & 2.775 & --- & --- & --- & 7.2 $\pm$ 0.2\\
		\hline
	\end{tabular}
        \\ {Column (1): Extraction aperture as defined in
          Fig.\ \ref{fig:ngc3079_apertures}; Columns (2) $-$ (3): Flux
          density at 100 and 160 $\mu$m in Jansky from the PACS data;
          Columns (4) $-$ (6): Flux density at 250, 350, and 500
          $\mu$m in Jansky from the SPIRE data; Column (7): Best-fit
          dust temperature in K; Column (8): Dust mass in solar masses
          derived from the best fit to the flux densities. The
          uncertainties on these quantities reflect the measurement
          errors, not the (possibly larger) systematic errors. See
          text for more details.}
\end{table*}

\begin{figure}
\centering
\includegraphics[width=0.45\textwidth]{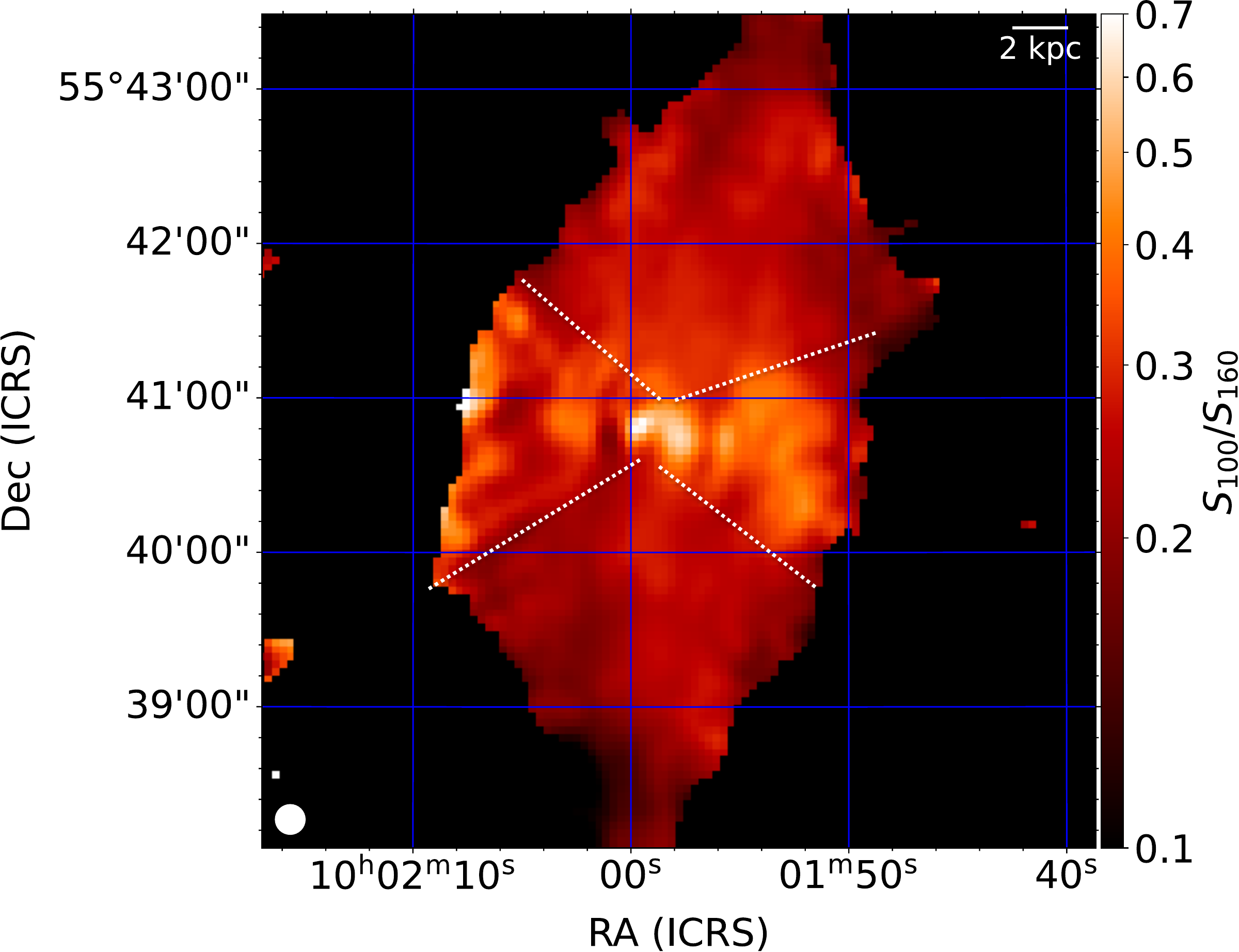} 
\caption{Ratio of the cleaned PACS 100 $\mu$m flux map to the cleaned
  PACS 160 $\mu$m flux map from Figure \ref{fig:ngc3079_pacs}. North
  is up and East to the left. The beam FWHM is displayed in the lower
  left corner, while the linear scale is shown in the upper right
  corner. The beam size is displayed in the lower left corner. The
  ratios are elevated in the biconical region indicated by the white dotted
  lines.}
\label{fig:ngc3079_pacs_100_160_ratio}
\end{figure}

The cleaned 100 and 160 $\mu$m flux maps of Figure
\ref{fig:ngc3079_pacs} were used to create the map of the
$S_{100}/S_{160}$ flux ratios presented in Figure
\ref{fig:ngc3079_pacs_100_160_ratio}, after convolving the 100 $\mu$m
flux map to match the beam size of the 160 $\mu$m flux map. The
$S_{100}/S_{160}$ ratios within the brighter portion of the galaxy
peak in the nucleus and within a bicone aligned along the minor axis
of the galaxy, coincident with the brightest X-ray emission
\citep[e.g.,][]{Strickland2004a}, the double-lobed radio structure
\citep[e.g.,][]{Irwin1988}, and the H$\alpha$ superbubbles
\citep[e.g.,][]{Veilleux1994}. This is similar to the results
found in both NGC~4631 (Paper I) and NGC~891 (Paper III), where the
$S_{70}/S_{160}$ ratios were found to be elevated within the X-ray
brightest regions in and above the nucleus of each galaxy. It is
  also reminischent of the ``ionization cones'' seen in starburst and
  active galaxies \citep[e.g.,][and references
    therein]{Shopbell1998,Wilson1994,Sharp2010}. We return to this
result in Section \ref{sec:discussion_nuclear}.

\subsection{PACS Spectroscopy}
\label{sec:results_imaging_spectroscopy}

\subsubsection{[C~II] 158 $\mu$m}
\label{sec:results_imaging_spectroscopy_cii}

The low-excitation atomic line \cii samples photodissociation regions
(PDRs) at the interfaces between molecular, atomic, and ionized gas
phases in or near star-forming regions or AGN where intense
far-ultraviolet (FUV) radiation photodissociates CO, resulting in
bright emission of [O I] and [C II]
\citep{Tielens1985,Sternberg1995,Hollenbach1997}. \cii arises from
both ionized and neutral gas due to the low ionization potential (11.2
eV) needed to create C$^+$. The relatively high second ionization
potential (24.4 eV) needed to ionize C$^+$ means that C$^+$ is often
the dominant atomic species of carbon. \cii is the dominant coolant in
regions with densities \nH\,$\sim10-10^5$ cm$^{-3}$ and temperatures
$T\sim100-300$ K, and is the strongest emission line from cooler gas
($T<10^4$ K) in galaxies \citep{Carilli2013}.

\begin{figure*}
\centering
\includegraphics[width=0.8\textwidth]{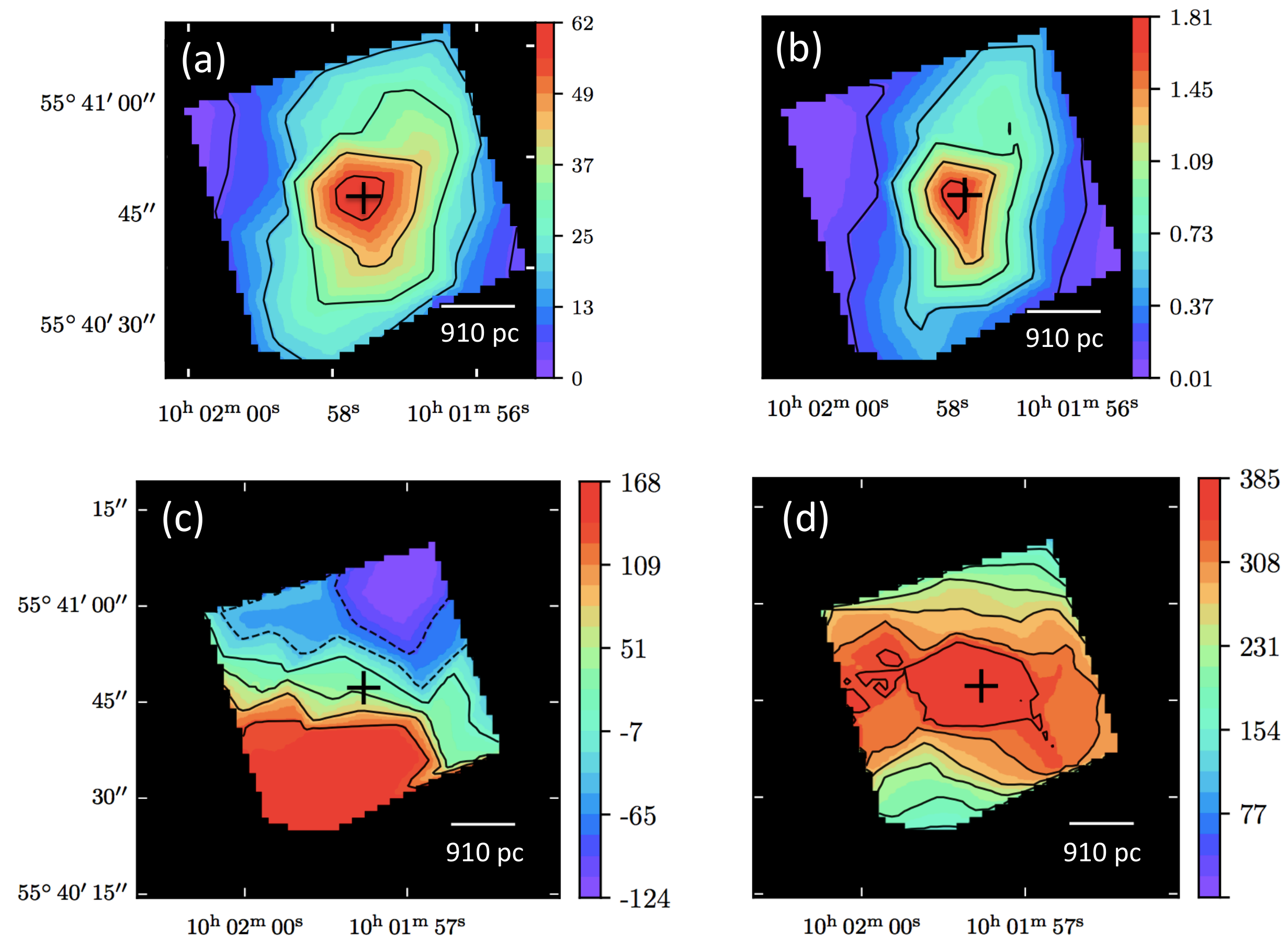}
\caption{(a) Signal-to-noise map of \cii in NGC~3079. Contours are
  0.1, 0.3, 0.5, 0.7, 0.8, and 0.9 of the peak value. The black cross
  in this and the other panels marks the adopted galaxy center. (b)
  Total integrated emission line fluxes in 10$^{-17}$ W
  m$^{-2}$. Contours are 0.1, 0.3, 0.5, 0.7, and 0.9 of the peak
  flux. (c) Map of the median velocities, \vFive, in \kms. Contours
  are in eight equal steps between the minimum and maximum
  velocities. (d) Map of the 1-$\sigma$ line widths, \wOne, in
  \kms. Contours are 0.4, 0.5, 0.6, 0.7, 0.8, and 0.9 of the peak
  width. North is up and East to the left. The linear
  scale is shown in each panel. }
\label{fig:ngc3079cii}
\end{figure*}

A signal-to-noise ratio map of the \cii line emission in NGC~3079 is
presented in Figure \ref{fig:ngc3079cii}a, where the largest SNRs are
observed in the nucleus with values exceeding $\sim$ 50. The
velocity-integrated \cii emission-line flux map is shown in Figure
\ref{fig:ngc3079cii}b. The \cii line emission in NGC 3079 is diffuse
and elongated along the direction of the major axis of the (inner)
disk, which runs along P.A. $\simeq$ 165$^\circ$.
Figures \ref{fig:ngc3079cii}c and \ref{fig:ngc3079cii}d present the
kinematics of the [C~II] line-emitting gas.  The velocities and widths
of the \cii emission line profiles in each spaxel are described using
a non-parametric method based on interpercentile range measurements.
\vFive is the median velocity of the fitted emission line profile,
i.e. 50\% of the emission is produced at velocities below \vFive. Zero
velocity corresponds to the rest wavelength at the systemic velocity
of NGC~3079. \wOne\ is the width of the line profile within 1-$\sigma$
standard deviation of \vFive\ (i.e. encompassing $\sim$ 68\% of the
total line flux).  The values of \cii \vFive\ range from $\sim$ $-$120
to $\sim$ $+$160 \kms and indicate that the gas on the N side is
approaching us while the gas on the S side is receding from us,
consistent with disk rotation traced by the warm-ionized,
neutral-atomic, and molecular gas
\citep{Veilleux1999,Koda2002,Yamauchi2004,Shafi2015}. The line widths
peak in the center and along the kinematic minor axis of the disk
i.e.\ along the same direction as the ionized and relativistic outflow
in this system. A more detailed interpretation of these data is
postponed until Section \ref{sec:discussion_nuclear}, where the
effects of beam smearing and instrumental broadening are taken into
account.

\subsubsection{OH 119 $\mu$m}
\label{sec:results_imaging_spectroscopy_oh}

We focus our attention on the ground-state OH 119 $\mu$m $^2 \Pi_{3/2}
J$ = 5/2 $-$ 3/2 rotation $\Lambda$-doublet. This feature is the
strongest transition in NGC~3079 and is positioned near the peak
spectroscopy sensitivity of PACS.
OH 119 is a sensitive tracer of the molecular gas. It is mainly
excited through absorption of far-IR photons and selectively traces a
region close to the central source of strong far-IR radiation density.

\begin{figure*}
\centering
\includegraphics[width=\textwidth]{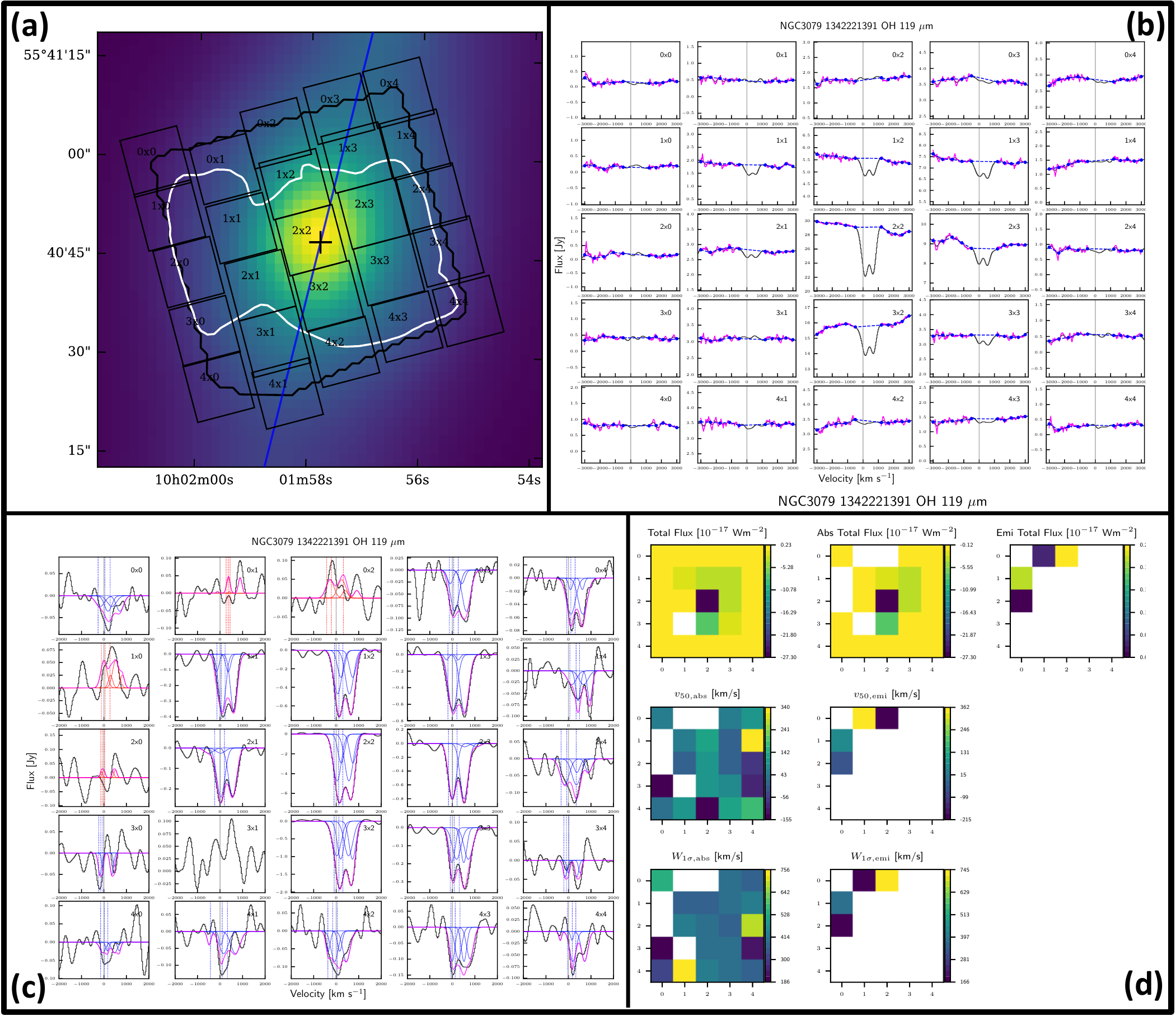}
\caption{ (a) PACS IFU footprint of OH 119 (black squares) overlaid on
  the 22 \um\,WISE image of NGC~3079. North is up and East to the left
  for this and all other panels in this figure. The white contour
  marks the spatial extent of the high-\wOne \cii region (defined in
  Section \ref{sec:definingWind}). The black contour outlines the PACS
  footprint of the \cii observation. The black cross marks the adopted
  galaxy center and the blue line marks the galaxy major axis. (b)
  Spline fits to the OH 119 continuum (blue dashed lines). Black lines
  are the observed data. Magenta areas indicate the regions used to
  fit the continuum. Blue dots mark the pivot points used to fit the
  spline. (c) Line profile fitting results of the continuum-subtracted
  spectra. Solid blue (red) lines indicate Gaussian absorption
  (emission) components. Vertical dashed blue (red) lines mark the
  \vSix, \vFive, and \vEight velocities in absorption (emission). (d)
  Top row, from left to right shows the total velocity-integrated flux
  of the fitted OH line profiles, the total flux in the absorption
  components only, and the total flux in the emission components
  only. Middle row shows \vFive for the absorption (left) and emission
  components (right). Bottom row shows the 1$-\sigma$ line widths of
  the absorption (left) and emission (right) components. OH line
  emission is not securely detected in this galaxy.}
\label{fig:ngc3079oh}
\end{figure*}

We characterize the OH line profiles in the same manner as the [C~II]
emission line (i.e.\ using \vFive and \wOne). The results are
summarized in Figure \ref{fig:ngc3079oh}.  Panel (a) of this figure
shows the PACS IFU footprint (black lines and squares) and the outline
of the region characterized with anomalously high \cii line widths
(white line; discussed in Sec.\ \ref{sec:discussion_nuclear}) overlaid
on the 22 $\mu$m WISE image of NGC~3079. Panel (b) shows the spline
fits to the continuum in each spaxel, all on the same flux scale,
while panel (c) presents the fits to the continuum-subtracted line
profile, adjusted to best show the results. Finally, panel (d) shows
maps of the line profile properties for the absorption and/or emission
components fitted to the spectra. The 1-$\sigma$ line width maps have
not been corrected for instrumental broadening.

It is clear from panels (b) and (d) in Figure \ref{fig:ngc3079oh} that
OH is largely seen only in absorption and is concentrated in the inner
3 $\times$ 3 $\times$ 9\farcs4 grid of the data cube, or $\sim$ 3
$\times$ 3 kpc centered on the nucleus. The OH line emission is
  not securely detected in this galaxy, thus the corresponding
  Gaussian fits shown in panels (c) and (d) of Figure
  \ref{fig:ngc3079oh} are unconstrained.  The N $-$ S 280 \kms
velocity gradient traced by \cii is not visible within the inner 3
$\times$ 3 grid where OH absorption is strongly detected. The OH
profiles in this region are uniformly broad with \wOne $\simeq$ 350
$-$ 450 \kms. These widths are similar to those of the [C~II] emission
line profiles in the same region. We return to these results in
Section \ref{sec:discussion_nuclear}.

\section{Discussion}
\label{sec:discussion}

\subsection{X-Shape Dust Structure}
\label{sec:discussion_large_scale}

\subsubsection{Dust Masses}

The dust masses were estimated by assuming a simple single-temperature
modified black body (MBB) to the infrared SED using
\begin{eqnarray}
F_\nu = \frac{M_d \kappa_\nu B_\nu(T_d)}{d^2}.
\label{eq:F_nu}
\end{eqnarray}
Here $M_d$ is the dust mass, $B_\nu$ is the Planck function, $T_d$ is
the dust temperature, $d$ is the distance to the galaxy, and
$\kappa_\nu$ is the dust emissivity, $\kappa_\nu = \kappa_0
(\nu/\nu_0)^\beta$, where $\kappa_0$ is the dust opacity at 350
$\mu$m. We follow the same procedure as in Papers I-III, choosing
$\kappa_0 = 0.192$ m$^2$ kg$^{-1}$ and $\beta$ = 2, and leaving the
dust temperature $T_d$ as a free parameter. This value of the dust
opacity is based on the best fit to the average far-infrared dust
emissivity for the Milky Way model presented in \citet{Draine2003},
which yields a best-fit spectral index value of $\beta$ = 2.0.



The resulting dust masses and temperatures are listed in Table
\ref{tab:dust_mass}. The detection of extraplanar emission at 250
$\mu$m in apertures A-E is uncertain so the dust masses derived within
these apertures, 1.2 $\times$ 10$^7$ $M_\odot$, should be considered
upper limits. The dust mass in aperture F, where extraplanar 250
$\mu$m emission is securely detected, provides a robust lower limit of
$\sim$ 4 $\times$ 10$^6$ $M_\odot$ for the total dust mass in the
X-shape structure. In the folowing discussion, we use (4 $-$ 16)
$\times$ 10$^6$ $M_\odot$ as the most plausible range of values for
the dust mass in the overall extraplanar structure.


\begin{figure}
\centering
\includegraphics[width=0.5\textwidth]{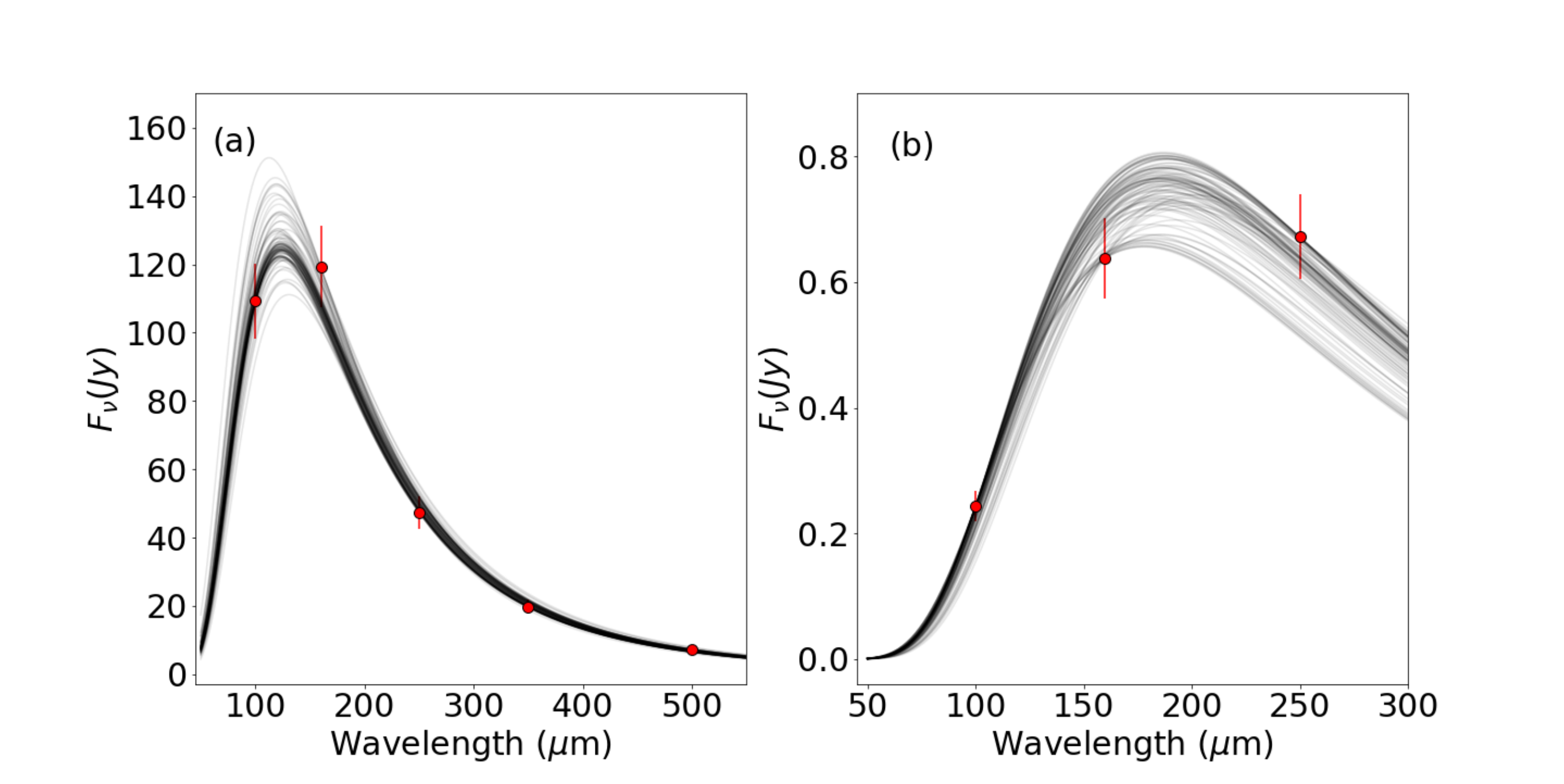}
\caption{Fits to the far-infrared energy distribution of (a) the
  global fluxes from NGC~3079, (b) the fluxes measured in Aperture
  F as defined in the top panel of
  Fig.\ \ref{fig:ngc3079_apertures}. }
\label{fig:ngc3079_sed_fit}
\end{figure}

\subsubsection{Multiwavelength Comparisons}

\begin{figure*}
\centering
\includegraphics[width=\textwidth]{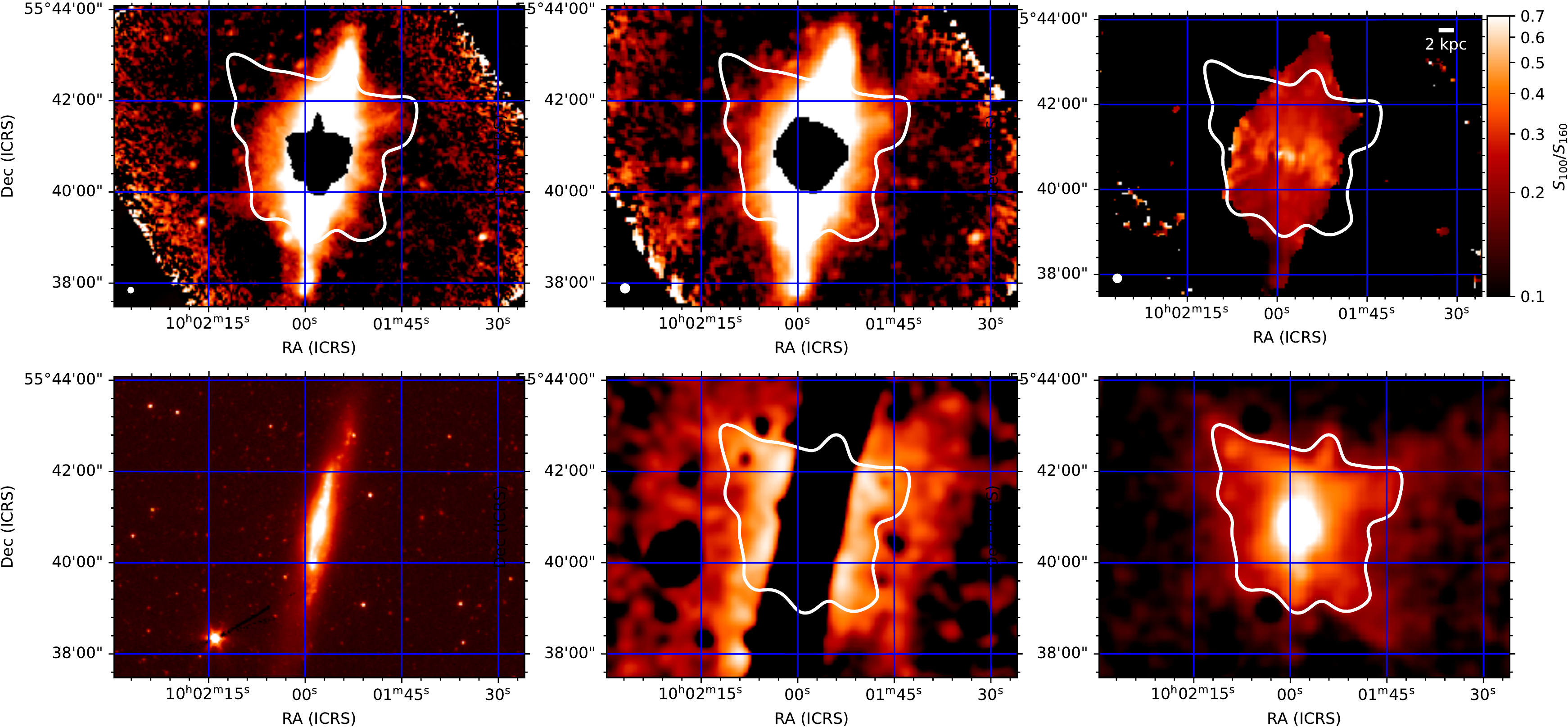}
\caption{Multi-wavelength comparison. (top left) PACS 100 $\mu$m
  map. (top center) PACS 160 $\mu$m map. (top right) PACS 100-to-160
  $\mu$m flux ratio map. The beam FWHM is displayed in the lower left
  corner of each panel. (bottom left) IRACS 4.5 $\mu$m image, (bottom
  center) FUV map from \citet{Hodges-Kluck2020}, (bottom right) {\em
    XMM-Newton} X-ray map from \citet{Hodges-Kluck2020}. The white
  contour in the various panels shows one of the X-ray isophotes for
  comparison. North is up and East to the left. The linear scale is
  shown in the upper right panel. All panels are on an
  arcsinh intensity scale. }
\label{fig:PACS_Xray_FUV_comp}
\end{figure*}

\begin{figure*}
\centering
\includegraphics[width=\textwidth]{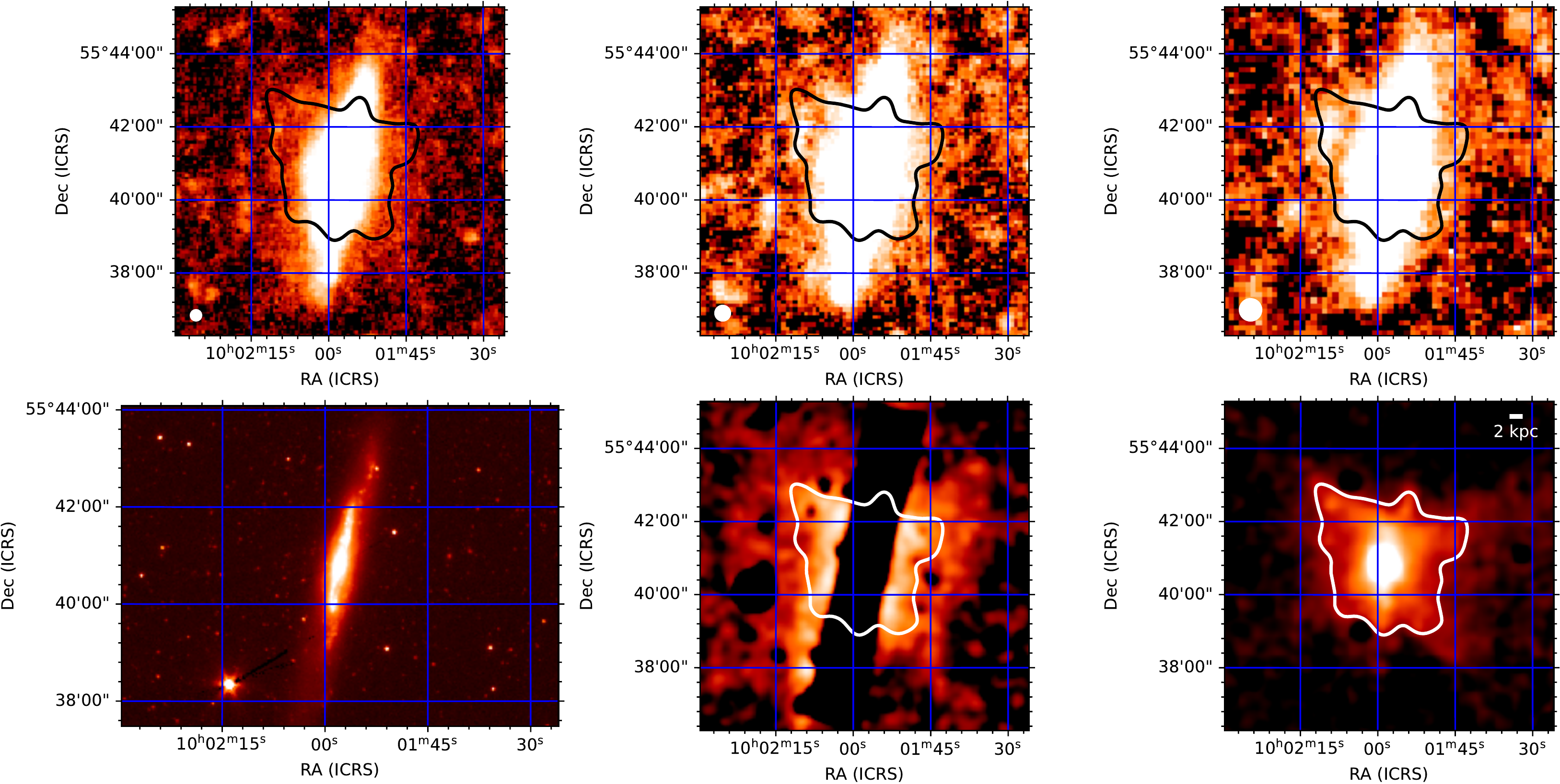}
\caption{Multi-wavelength comparison. (top left) SPIRE 250 $\mu$m
  map. (top center) SPIRE 350 $\mu$m map. (top right) SPIRE 500 $\mu$m
  map. The beam FWHM is displayed in the lower left corner of each
  panel. (bottom left) IRACS 4.5 $\mu$m image, (bottom center) FUV map
  from \citet{Hodges-Kluck2020}, (bottom right) {\em XMM-Newton} X-ray
  map from \citet{Hodges-Kluck2020}. North is up and East to the
  left. The white contour in the various panels shows one of the X-ray
  isophotes for comparison. North is up and East to the left. The
  linear scale is shown in the lower right panel. All panels are on an
  arcsinh intensity scale. }
\label{fig:SPIRE_Xray_FUV_comp}
\end{figure*}

\begin{figure*}
\centering
\includegraphics[width=\textwidth]{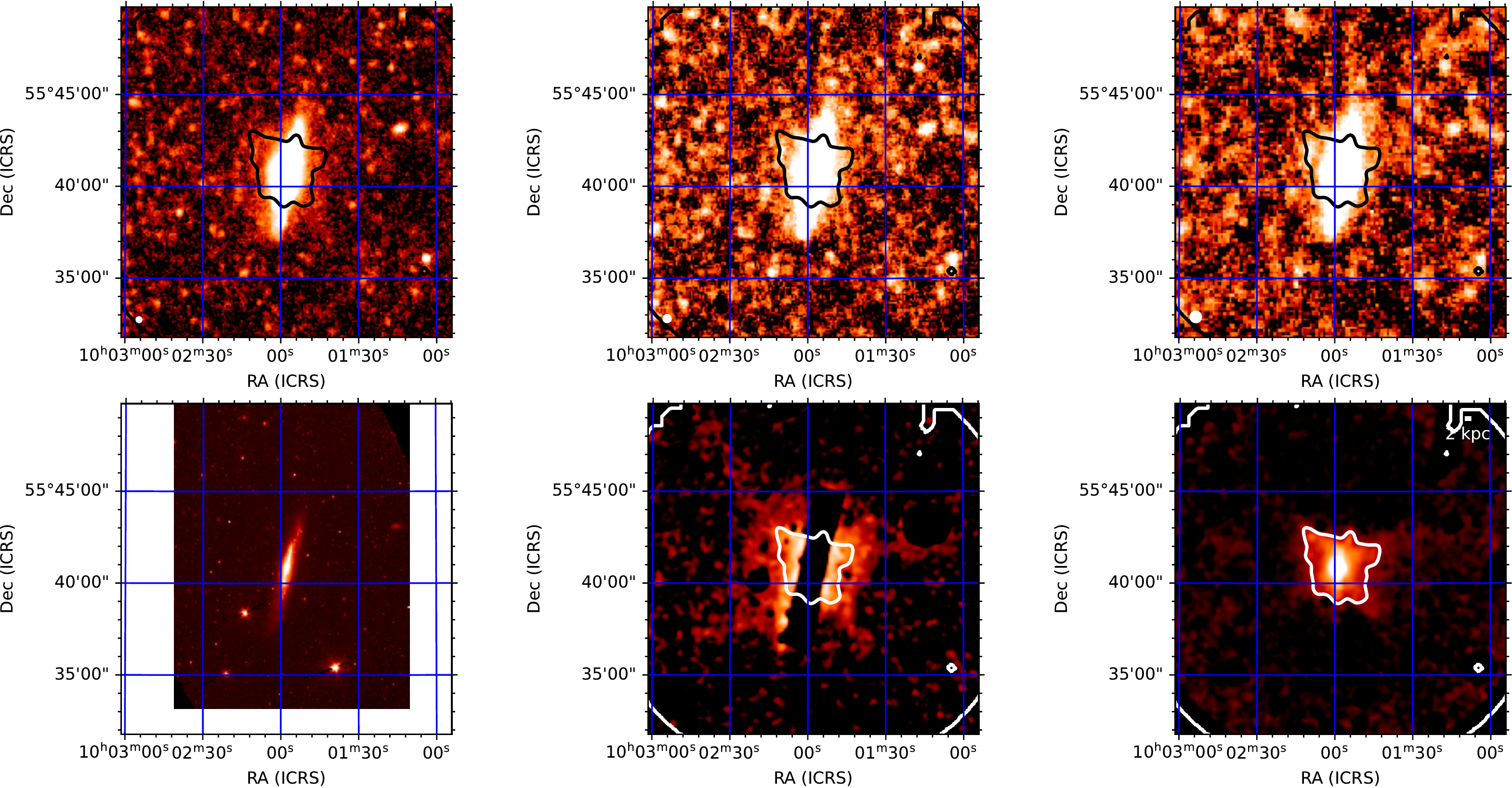}
\caption{Same as Figure \ref{fig:SPIRE_Xray_FUV_comp} but showing a
  larger field of view. (top left) SPIRE 250 $\mu$m map. (top center)
  SPIRE 350 $\mu$m map. (top right) SPIRE 500 $\mu$m map. The beam
  FWHM is displayed in the lower left corner of each panel. (bottom
  left) IRACS 4.5 $\mu$m image, (bottom center) GALEX FUV map from
  \citet{Hodges-Kluck2020}, (bottom right) {\em XMM-Newton} X-ray map
  from \citet{Hodges-Kluck2020}.  The white contour in the various
  panels shows one of the X-ray isophotes for comparison. North is up
  and East to the left. The linear scale is shown in the lower right
  panel. All panels are on an arcsinh intensity scale. }
\label{fig:SPIRE_Xray_FUV_comp_largerFOV}
\end{figure*}

Figures \ref{fig:PACS_Xray_FUV_comp}, \ref{fig:SPIRE_Xray_FUV_comp},
and \ref{fig:SPIRE_Xray_FUV_comp_largerFOV} compare the far-infrared
PACS and SPIRE emission of NGC~3079 with the deep {\em GALEX} and {\em
  XMM-Newton} data recently analyzed by
\citet{Hodges-Kluck2020}. There is an excellent match between the
far-infrared X-shape structure and the base of the biconical outflow
visible in the far-ultraviolet and soft X-ray maps. This X-shape
structure has also been detected at H$\alpha$ \citep[Figs.\ 7 and 5
  in][respectively]{Heckman1990,Veilleux1995}. These results suggest
  that the far-infrared X-shape structure is physically associated
  with the large-scale galactic wind of NGC~3079. In this picture,
  dusty material originally in the galactic disk of NGC~3079 is lifted
  above the disk and entrained in the large-scale galactic wind.

\subsubsection{Implications on the Energetics}
\label{sec:energetics}

The energy needed to lift the dusty material making up the X-shape
structure may be derived following the same procedure as that used in
Papers I and III. First, we need to estimate the total (gas + dust)
mass associated with the extraplanar dust. Here, it is important to
account for all possible gas phases, which is challenging. The amount
of hot ionized material associated with the base of the X-ray emitting
filaments, $\sim$ 4 $\times$ 10$^7$ $M_\odot$ \citep[from Table 1
  of][excluding the central region]{Hodges-Kluck2020}, implies a ratio
of the hot ionized gas mass to the dust mass of 3 $-$ 10, which is a
hard lower limit on the total gas-to-dust mass ratio. On the other
hand, the amount of warm ionized gas in the X-shape H$\alpha$
structure remains unknown; H$\alpha$ is not only very faint but the
large uncertainties on the 3D geometry on the H$\alpha$ filaments
(e.g., cylinders {\em versus} conical walls in projection) make it
difficult to derive a reliable gas density in this structure from the
H$\alpha$ emission (typical line ratio diagnostics such as the [S~II]
$\lambda$6731/$\lambda$6716 cannot be used in this low-density
regime). While cooler gas no doubt also contributes to increasing the
gas-to-dust mass ratio, there is no report at present of extraplanar
molecular gas on the scale of the X-shape structure
\citep[e.g.,][]{Sofue1992}. Finally, extraplanar H~I gas has been
detected in NGC~3079 \citep[see Fig.\ 1 in][]{Shafi2015}, but there is
no obvious overlap between the H~I emission and the X-shape structure
down to a sensitivity level corresponding to an H~I column density of
$\sim$ 0.05 $\times$ 10$^{20}$ cm$^{-2}$.

In the end, given the large uncertainties on the derivation of the
gas-to-dust mass ratio in the X-shape structure, we make the
simplifying assumption that the gas-to-dust ratio of the X-shape
structure is similar to that of the disk ISM, i.e. it has not been
modified by the entrainment in the wind. We return to the validity of
this assumption below (Sec.\ \ref{sec:dust_entrainment}). In that
case, a Galactic dust-to-gas mass ratio of 100 $-$ 150
\citep[e.g.,][]{Draine2007} implies a gas mass of (4 $-$ 24) $\times$
10$^8$ $M_\odot$. On the other hand, \citet{Yamagishi2010} derived a
gas-to-dust mass ratio of $\sim$ 860 for NGC~3079 as a whole and
$\sim$ 1100 within the central (2 kpc radius) region. If these
unusually high gas-to-dust mass ratios also apply to the extraplanar
material, then the gas mass in the X-shape structure would be up to (3
$-$ 18) $\times$ 10$^9$ $M_\odot$.

We have reexamined this question using our {\em Herschel} data and
fitting the global far-infrared SED of NGC~3079 with a
single-temperature MBB, $\kappa_0$ = 0.192 m$^2$ kg$^{-1}$, and
$\beta$ = 2, as we did for the X-shape structure. The results are
listed on the first line of Table 1 and shown in the left panel of
Figure \ref{fig:ngc3079_sed_fit}. The flux measurements are consistent
with other published values from shallower data
\citep{Melendez2014,Shimizu2016}.  The derived total dust mass of $1.0
\times 10^8$ $M_\odot$ is $\sim$ 5 $\times$ higher than the value
adopted by \citep[][scaled to our adopted distance of 19
  Mpc]{Yamagishi2010}. The global gas-to-dust mass ratio we derive
from our data, using an updated total (molecular + atomic) gas mass of 2
$\times$ 10$^{10}$ $M_\odot$ \citep[combining the values
  from][]{Devereux1990,Sofue1992,Irwin1987,Shafi2015},
is 200. Using this range of
gas-to-dust ratios for our calculation of the gas mass in the X-shape
structure, we get (8 $-$ 32) $\times$ 10$^8$ $M_\odot$. Interestingly,
note also that the best-fit dust temperature (24 K) to the global SED
is higher than that of the X-shape filaments. We return to this issue
below (Sec.\ \ref{sec:dust_entrainment}).


Following Papers I and III, we estimate the energy needed to lift this
material up to a height $z$ above the mid-plane of the galaxy by
assuming an isothermal sheet model for the vertical distribution of
light and mass in galactic disks \citep{Howk1997}:


\begin{eqnarray}
\Omega & = & 1 \times 10^{56}~{\rm ergs}\left ( \frac{M}{10^9
  M_\odot} \right ) \left( \frac{z_o}{700~{\rm pc}} \right )^2 \left
(\frac{\rho_o}{0.185~{\rm M_\odot~pc^{-3}}}\right ) \nonumber \\ & &
\times \ln \left [\cosh \left ( \frac{z}{z_o}\right ) \right ],
\label{PE}
\end{eqnarray} 

\noindent where $\rho_o$ is the mass density at the midplane, $M$ is
the mass of lifted material, and $z_o$ is the mass scale height of the
stellar disk. Here we use a mass scale height for the stellar disk of
380 pc, derived at $\sim$ 2 \micron\ \citep{Veilleux1999}, and ignore
the well-known boxy, peanut-shaped bulge in this object \citep[][and
  references therein]{Veilleux1999} since we are only concerned with
the large-scale X-shape filaments which are not affected by the inner
bulge.  Assuming an average mid-height for the X-shape structure above
the plane of about 3~kpc, and a starting point at $z$ = $z_0$ = 380
pc, then the gain in potential energy is:

\begin{equation}
\Delta\Omega = 1\times 10^{56}~{\rm ergs} \left ( \frac{M}{10^9
  M_\odot} \right )\left (\frac{\rho_o}{0.097~{\rm
    M_\odot~pc^{-3}}}\right ).
\label{PE1}
\end{equation} 

Note the rather weak dependence on the starting point: if the material
starts at a distance ten times closer to the mid-plane, $z$ = 0.1
$z_0$ = 38 pc, then the energy needed to lift the material would be
only $\sim$ 1.05 $\times$ higher (note that eqn~\ref{PE} is undefined
at the mid-plane). The same cannot be said about the average
mid-height for the X-shape structure: a value of 1 kpc instead of 3
kpc gives $\Delta\Omega$ = 0.13 $\times$ 10$^{56}$ ergs. Another
uncertainty on the prediction of the potential energy comes from the
mass density at mid-plane, where we adopted a value of 0.097 ${\rm
  M_\odot~pc^{-3}}$, representative of the total mass density at the
solar position \citep[see][for details]{Bland-Hawthorn2016}. This is
probably a good first-order approximation since NGC~3079 and the Milky
Way share similar morphological, kinematic, and dynamical properties
\citep{Veilleux1999}.

This gain in potential energy is large, the equivalent of about 10$^5$
supernova explosions, and larger than the kinetic energy of the
nuclear H$\alpha$ line-emitting bubble, (0.5 $-$ 6) $\times$ 10$^{55}$
$\sqrt{f}$ ergs, regardless of the gas filling factor $f$ \citep[$<$
  1;][for $d$ = 19 Mpc]{Cecil2001}. This last comparison is not
completely fair since the nuclear bubble extends only on kpc scale
while the infrared filaments reach distances of up to 25 kpc. The
dynamical time scale of the nuclear bubble is $\sim$ 10$^6$ yrs
\citep{Cecil2001}, while the filaments were likely produced over a
much longer time scale. For instance, the lifetime of the 60-kpc
galactic wind reported in \citet{Hodges-Kluck2020} is estimated to be
120 Myr, two orders of magnitude larger than the bubble time
scale. Using a AGN jet power of $\sim$ 6 $\times$ 10$^{41}$
erg~s$^{-1}$ from \citet{Shafi2015}, adjusted for $d$ = 19 Mpc, which
accounts for the various methods that can be used to convert radio
luminosities into jet power, the energy needed to lift the X-shape
structure would require the jet to inject 50\% of its kinetic energy
and remain turned on for $\sim$ 10$^7$ years. This time scale is long
but not unreasonable: for instance, the buoyancy time scale of the
radio structure in NGC~3079 has been estimated to be also $\sim$
10$^7$ yr \citep{Cecil2001}.

In this mass-loaded AGN-driven wind scenario, the AGN injects
  material in the ISM at a rate of $\sim L_{\rm
    kin}/(\frac{1}{2}~V_{\rm wind}^2)$, where $V_{\rm wind}$ is the
  AGN wind velocity at the source and the AGN wind kinetic power
  $L_{\rm kin}$ needs to average out to at least $\sim$ 50\% of the
  AGN jet power, or $\sim 3 \times 10^{41}$ erg~s$^{-1}$, for a period
  of $\sim$ 10$^7$ years to be able to lift the dusty material into
  the halo. This mass outflow rate is only 0.01 $-$ 1 $M_\odot$
  yr$^{-1}$ for $V_{\rm wind}$ = 1000 $-$ 10,000 \kms, more than 2-4
  orders of magnitude smaller than the time-averaged mass ejection
  rate derived by dividing the inferred gas mass in the X-shape
  structure by $10^7$ yrs.  In the pure-starburst scenario, which we
  consider less likely, the starburst injects material in the ISM at a
  rate $\sim$ 0.26 SFR \citep{Veilleux2005}, where SFR $<$ 2.6
  $M_\odot$ yr$^{-1}$ \citep{Yamagishi2010}, or about two orders of
  magnitude smaller than the above time-averaged mass ejection rate
  into the halo. Thus, regardless of the nature of the wind energy
  source (AGN or starburst), the implied large mass-loading factor
  (defined as the rate of mass ejected into the halo divided by the
  mass outflow rate produced at the source) adds support to the idea
  that most of the dust in the halo originates from disk material
  being mass-loaded into the wind. Taken at face value, these
  mass-loading factors are considerably larger than those predicted by
  numerical simulations of isolated galaxies \citep{Hopkins2012}.
  However, caution should be exercised when interpreting these results
  since the derived mass-loading factors are only order-of-magnitude
  estimates and the simulations are idealistic in the sense that they
  do not include a realistic hot coronal gas component or
  intergalactic medium.

So far, our discussion has focused on the necessary (minimum)
condition of lifting the material into the halo. In addition, we also
need to take into account the work done against the surrounding ISM
and CGM.  A shell that moves faster than the sound speed will
accelerate as it breaks through the disk. Rayleigh-Taylor
instabilities will disrupt the shell and a free-flowing galactic wind
will develop \citep{Chevalier1985}. The critical rate of energy
injection for blowout to occur depends on the vertical gas density
distribution. Using equation (5) of \citet{Strickland2004b} (or,
equivalently, equation (2) in Paper III), the critical mechanical
power to break out through the thick H~I disk of NGC~3079 is
\begin{equation}
L_{\rm crit}= 10^{38}~{\rm ergs~s}^{-1} \left ( \frac{0.1~{\rm cm}^{-3}}{n_0} \right )^{1/2}\left (\frac{P_0/k}{10^4~{\rm K~cm}^{-3}}\right )^{3/2}\left (\frac{H}{\rm kpc}\right )^2
\label{eq:Lcrit}
\end{equation} 
where we have assumed that the vertical gas density profile of the
thick disk of NGC~3079 follows approximately an exponential
distribution with a scale height $H$ = 1 kpc and central gas density
$n_0$ = 0.1 cm$^{-3}$ \citep[][using $d$ = 19 Mpc]{Irwin1991}. The
rate of energy injection from the AGN jet in NGC~3079 ($\sim$ 6
$\times$ 10$^{41}$ erg~s$^{-1}$) exceeds this breakout criterion by a
factor $>$1000. The energy injection rate also exceeds the critical
rate to break out of the thin molecular disk
\citep{Sofue2001,Koda2002} and thick warm-ionized gas layer
\citep{Veilleux1995}.

This picture of a powerful, long-lived AGN-driven outflow that
breaks out from the disk can also naturally explain the large-scale
X-shape morphology seen in the far-infrared, UV, and X-rays. The
breakout flow is expected to be a strong function of polar angle
as seen from the nucleus \citep{Cecil2001}. Neglecting the CGM, the
flow expands freely along the minor axis, but at some critical angle
its ram pressure becomes comparable to the thermal pressure in the
diffuse component of the ISM; at larger angles a standing bow shock in
the galaxy disk decelerates and deflects wind around a region of
undisturbed ISM gas. The wind subsequently reaccelerates into the
halo. In the disk plane itself, the flow compresses the ISM to form a
standing ``ring'' shock at $r \simeq$ 800 pc. \citet{Cecil2001} have
argued that the linear X-shape H$\alpha$ filaments join the galaxy
disk at radii of $\sim$ 800 pc, forming a ``concave bowl'' in the {\em
  HST} images that coincides with the standoff outer disk shock /
contact-discontinuity between the shocked wind and the unperturbed ISM
(see their Figs.\ 10 and 12 for instance). Unfortunately, the limited
angular resolution of the {\em Herschel} data, combined with the
uncertainties associated with removing the PSF from the PACS images,
prevent us from tracing the X-filaments of the far-infrared structure
inside of $\sim$ 2-3 kpc and determining whether the match between the
far-infrared and H$\alpha$ filaments continues down to the nuclear
scale.

The actual situation in NGC~3079 is undoubtedly more complicated:
interaction with the CGM will slow down the free-flowing wind, perhaps
allowing it to cool and form a galactic fountain rather than escape
the gravitational potential. Equation \ref{eq:Lcrit} cannot be used to
estimate the critical mechanical power needed to break out through the
CGM as it neglects gravity, radiative losses, and the density profile
of the CGM \citep[see][for examples of semi-analytic models of
  starburst-driven winds]{Lochhaas2018}.



\subsubsection{Dust Entrainment and Survival}
\label{sec:dust_entrainment}

Several processes may destroy the dust entrained in the outflow
\citep[e.g.,][]{Galliano2018,Veilleux2020}: collisions with other
grains, sputtering due to collisions with ions, sublimation or
evaporation, explosion due to ultraviolet radiation, and alteration of
grain material by cosmic rays and X-rays.  A distinction is often made
between thermal sputtering, where the sputtering rate only depends on
the local gas properties (namely $n_H$ and $T$), and non-thermal
(inertial) sputtering, where the dust-gas relative velocity is a
crucial parameter.
The thermal sputtering timescale is
\begin{eqnarray}
t_{\rm sput} \simeq 3.3 \times 10^3~\left(\frac{a}{0.1~\mu{\rm
    m}}\right)~\left(\frac{n_H}{10~{\rm
    cm}^{-3}}\right)^{-1}~{\rm yr},
\label{eq:t_sput}
\end{eqnarray}
for $10^6$ K $\lesssim$ $T$ $\lesssim$ 10$^9$ K, and increases
dramatically below $\sim$10$^6$ K \citep{Nozawa2006,Hu2019}. Assuming
the hot ($T \simeq 0.35$ keV = 4 $\times$ 10$^6$ K), low-density ($n_H
\simeq 10^{-3}$ cm$^{-3}$), X-ray emitting material in the X-shape
filaments reported by \citet{Hodges-Kluck2020} is in contact with the
dust would imply a thermal sputtering timescale of 3 $\times$ 10$^7$
yrs for dust grains with $a = 0.1$ $\mu$m, similar to the duration of
the AGN jet cycle needed to produce the X-shape structure.

The presence of dust in the X-shape filaments would seem to exclude
shocks faster than $\sim$ 300 \kms, which would rapidly destroy the
dust through non-thermal (inertial) sputtering
\cite[e.g.,][]{Arendt2010,Lakicevic2015,Temim2015,Dopita2016},
although the presence of dust behind the reverse shock in some
supernova remnants \citep[e.g.,][]{Kochanek2011,Matsuura2019} and in
galaxies in general \citep{Gall2018} suggests that dust is either more
resilient than originally predicted
\citep[e.g.,][]{Silvia2010,Biscaro2016} or it reforms rapidly behind
the shocks \citep{Humphreys2012,Seale2012,Gall2014}, despite models
predicting the contrary \citep[e.g.,][]{Biscaro2014}.

\citet{Hodges-Kluck2020} argued that shocks with velocities exceeding
500 km~s$^{-1}$ are needed to produce the large-scale X-ray filaments
if the X-rays come from cool material shock-heated by the wind. To
survive, the dust must be shielded from these fast shocks, and the low
dust temperature coincident with these filaments ($\sim$15~K, compared
to 25~K in the disk) indicates that the only heat source is radiation
leaking out of the disk. This is also consistent with the
interpretation that the extraplanar FUV and NUV light at low latitudes
is reflected by dust in these filaments, and indeed the diffuse UV
light is brightest in the same region where the X-ray filaments are
brightest. The presence of the dust raises the possibility that the
hot gas is not primarily shock heated, but instead traces disk
material mass-loaded into the wind that has mixed with a hotter phase
\citep[as envisioned in][]{Cooper2008}. The wind is sufficiently
mature that this is a strong possibility, as any terminal shock is at
least 60~kpc away. In this case, the density and temperature do not
map straightforwardly to the wind speed, and the thermal sputtering
timescale (or cloud evaporation timescale) is the most appropriate one
for grain survival.

\subsection{Cool Nuclear Outflow}
\label{sec:discussion_nuclear}

In Section \ref{sec:results_imaging_spectroscopy}, we showed that the
\cii velocity field of NGC~3079 is dominated by the rotational motion
of the gas in the disk, while that of the OH absorption feature is
not. However, in both cases, the profiles near the center are broad
(\wOne = 350 $-$ 450 \kms), suggestive of a turbulent medium. To more
quantitatively assess the influence of the nuclear outflow on the
kinematics of \cii line profiles, we need to remove the instrumental
effects associated with the finite spectral and spatial resolutions of
the data and examine the residuals. This is a multi-step process which
is detailed here.


\subsubsection{Beam Smearing}
\label{sec:beamSmearing}

An important aspect of deriving accurate galaxy kinematics from
velocity fields is to take into account the effect of beam smearing
\citep{Bosma1978}. The radial velocities in a galaxy vary on spatial scales
smaller than the beam size of the observing instrument.  Effectively,
this means that the velocities at different radii will be blended,
thus flattening the observed velocity field gradient, decreasing the
slope of the derived rotation curve at the galaxy center, and
broadening the width of the observed line profiles. This is a
significant problem since the broadening effect can be
incorrectly attributed to intrinsic gas velocity dispersion.
 
Fortunately, we can employ methods that utilize the full 3D data cube
(two spatial dimensions and one spectral dimension) of the galaxy and
build a model that directly incorporates the instrument 2D point
spread function and spectral resolution. To do this we use the
software package
\barolo\footnote{http://editeodoro.github.io/Bbarolo/, version 1.4}
\citep{DiTeodoro2015} which simulates the observational data
in a cube by building a ``tilted-ring'' model that best fits the
data. Accounting for the instrumental contribution to the observed
data will result in a disk model which more accurately captures the
true kinematics of the gas. The details of the ``tilted-ring'' model
and the employment of \barolo\ are discussed in the following section.

\subsubsection{Tilted Ring Model}
\label{sec:ringModel}

The velocity field of a disk galaxy can be described by a
``tilted-ring'' model \citep{Rogstad1974}, where the disk is built
from a series of concentric annuli of various radii.
This model assumes that the line emitting material is confined to a
thin disk and that the kinematics are dominated by rotational
motion. Emission of the gas in each ring is described by geometric
parameters (centroid, radius, width, scale height, inclination angle
along the line of sight of the observer, and position angle of the
major axis) and kinematic parameters (systemic velocity, rotational
velocity, and velocity dispersion). For \barolo, the instrumental
spectral and spatial resolutions are also inputs so that the final
model accounts for both instrumental effects.

In reproducing the observed data cube, \barolo\, assumes that all of
the velocity dispersion is due to rotation and turbulence. Given that
an outflow may exist along the minor axis of NGC~3079, this
assumption means that the velocity dispersion of the disk model will
be overestimated, with the largest errors at the center of the galaxy
where line broadening suffers the most from beam smearing.

We attempt to mitigate this effect by assuming a constant velocity
dispersion across the disk and estimate the value of this velocity
dispersion using data outside the galaxy center where beam
smearing is the least severe. The data are fit in two stages. In the
first stage, the gas velocity dispersion and the rotational velocity
are left as free parameters. The remaining parameters are input as
``correct'' values and held constant. Afterwards, the mean of the
fitted gas velocity dispersions of the three outermost rings is then
computed. For the second stage, this mean dispersion is input as a
fixed parameter and only the rotational velocity is left free. The
resultant disk model is the assumed galaxy velocity field.

\begin{figure*}
\centering
\includegraphics[width=\textwidth]{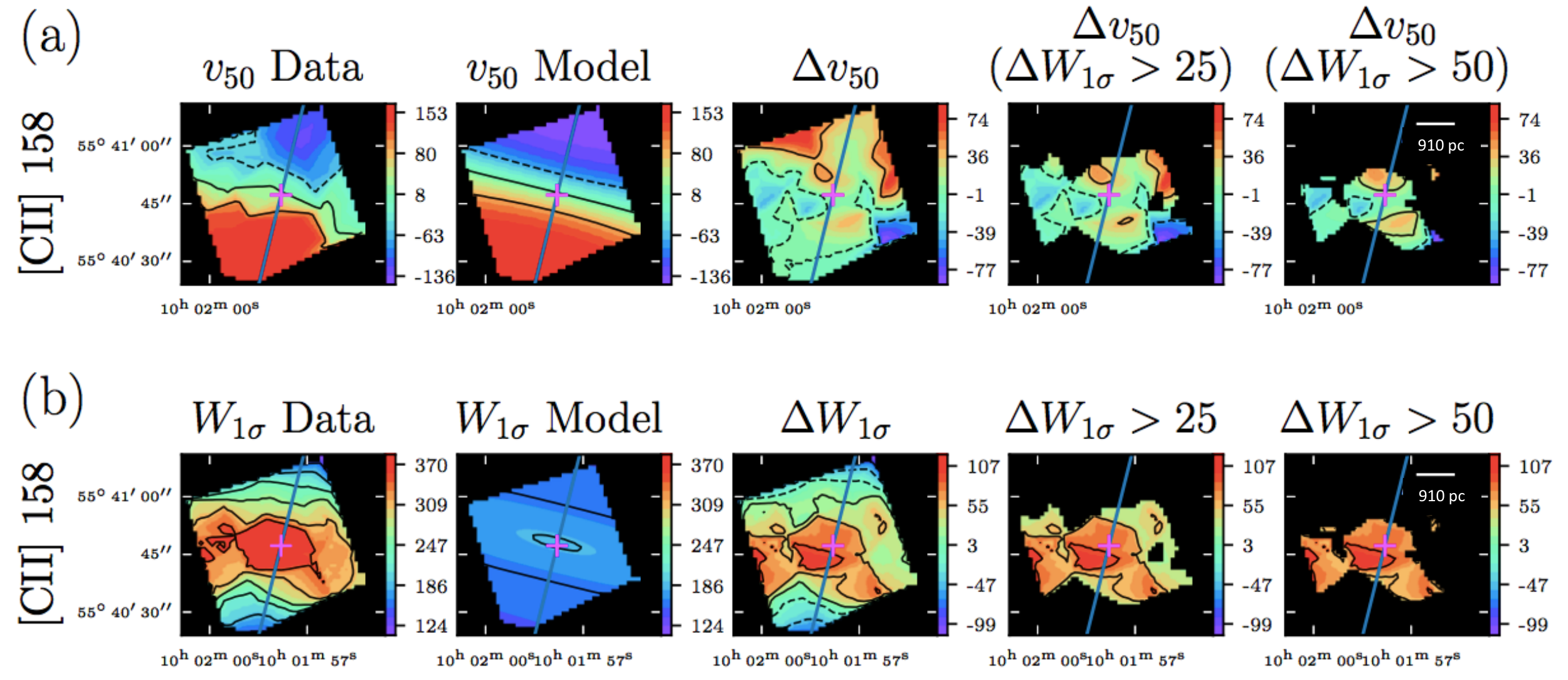}
\caption{Results from modeling the \cii disk velocity field (a) and
  line widths (b) of NGC~3079 with \barolo\,(which accounts for both
  the instrumental spectral and spatial resolutions), and the location
  of the outflow based on excess line broadening. The values of the color
  bar are in units of \kms. For each row of panels, we show from left to right
  the observed data, the best-fit model, the data $-$ model residuals,
  the spatial location of the wind in regions where \wOneDelt $>25$
  \kms, and the spatial location of the wind in regions where
  \wOneDelt $>50$ \kms. Contours in (a) are in five equal steps
  between the minimum and maximum velocities in each image. Contours
  in (b) are 0.3, 0.5, 0.7, and 0.9 of the peak value in each
  image. The solid blue line marks the galaxy major axis. The magenta
  cross marks the adopted galaxy center. North is up and East to the left.}
\label{fig:ngc3079_v50_w1}
\end{figure*}

\subsubsection{Defining the Impact of the Outflow}
\label{sec:definingWind}

Any potential impact of the outflow on the [C~II] kinematics may be
delineated from those of the disk by examining the residuals in
\vFive\,and in \wOne\,between the observed data and the modeled disk
(\vFiveDelt\ and \wOneDelt, respectively). Recall from Section
\ref{sec:ringModel} that the modeled galaxy velocity field accounts
for the instrumental spatial and spectral resolutions. Therefore, in
the residuals, the instrumental effects have been
removed. Experimentation on {\em Herschel} PACS data of other galaxies
with known spatially resolved outflows along their minor axes,
including M~82 \citep{Contursi2013}, has shown that the excess line
widths are a more reliable indicator of the outflow than the
\vFive\ residuals, which are more directly affected by common
dynamical features in disk galaxies, such as spiral arms, nuclear
bars, and warped disks \citep{Stone2020}.
This is also true for NGC~3079 which hosts a weak bar that influences
the kinematics of the gas in the central $\sim$ 10\arcsec\ region
\citep[e.g.,][]{Veilleux1999,Koda2002}.

While \wOneDelt\ is a more robust tracer of outflows than \vFiveDelt,
the impact on \wOneDelt\ from non-circular motions in the gas around
spiral arms, stellar bars, and disk warps cannot be ignored.
Therefore, when defining the spatial location of the wind, it is
important to explore different minimum thresholds in \wOneDelt which
would account for gas motions not related to the outflow.
The results for two thesholds, \wOneDelt\ $ >25$ \kms\ and
\wOneDelt\ $>50$ \kms, are shown in Figure \ref{fig:ngc3079_v50_w1}.
Our experimentation on other galaxies with known spatially resolved
outflows indicates that the smaller of these two velocity thresholds
already minimizes the inclusion of regions where the excess line
widths are spatially coincident with spiral arms, stellar bars, and
disk warps in the central regions where the outflows are located
\citep{Stone2020}.
While the higher velocity threshold does more effectively exclude gas
motions unrelated to the outflows, such a threshold has a tendency to
exclude a large portion of [C~II] outflows altogether, as in the case
of M~82 \citep{Contursi2013}. We therefore favor the lower of these
two line width thresholds although, as shown in Figure
\ref{fig:ngc3079_v50_w1}, the results of our analysis in NGC~3079 are
not sensitive to this threshold value.

\subsubsection{Wind Kinematics and Multiwavelength Comparison}
\label{sec:multiwavelength}

The \cii \vFive\ residuals in the wind region
(Fig.\ \ref{fig:ngc3079_v50_w1}a, Columns 4-5) do not show any regular
pattern.
In contrast, the \cii \wOneDelt\ map (Fig.\ \ref{fig:ngc3079_v50_w1}b,
Column 4-5) shows a clear E-W elongated feature with
\wOneDelt\ $\ga100$ \kms, about 10\arcsec\ in length and off-centered
to the SE from the nucleus. The contour marking \wOneDelt\ $\sim75$
\kms\ delineates a more extended E-W structure that encompasses this
elongated high-\wOneDelt\ feature.

A comparison of the \cii line profiles with those of CO (1-0) \citep[e.g.,
Figure 9 of][]{Koda2002}
indicates that the \cii line-emitting gas is largely unaffected by the
bar streaming and spiral arm disturbances detected in the kinematics
of the molecular gas. We see no obvious match in our data with the
double-peaked CO (1-0) profiles observed SE and NW of the nucleus in
the data of \citet{Koda2002}. Moreover, \cii \vFiveDelt\ and
\wOneDelt\ extend further ($>$10\arcsec) and along a different
orientation (E-W) than the CO (1-0) nuclear disk and core reported by
\citet{Koda2002}.

\begin{figure*}
\centering
\includegraphics[scale=.80]{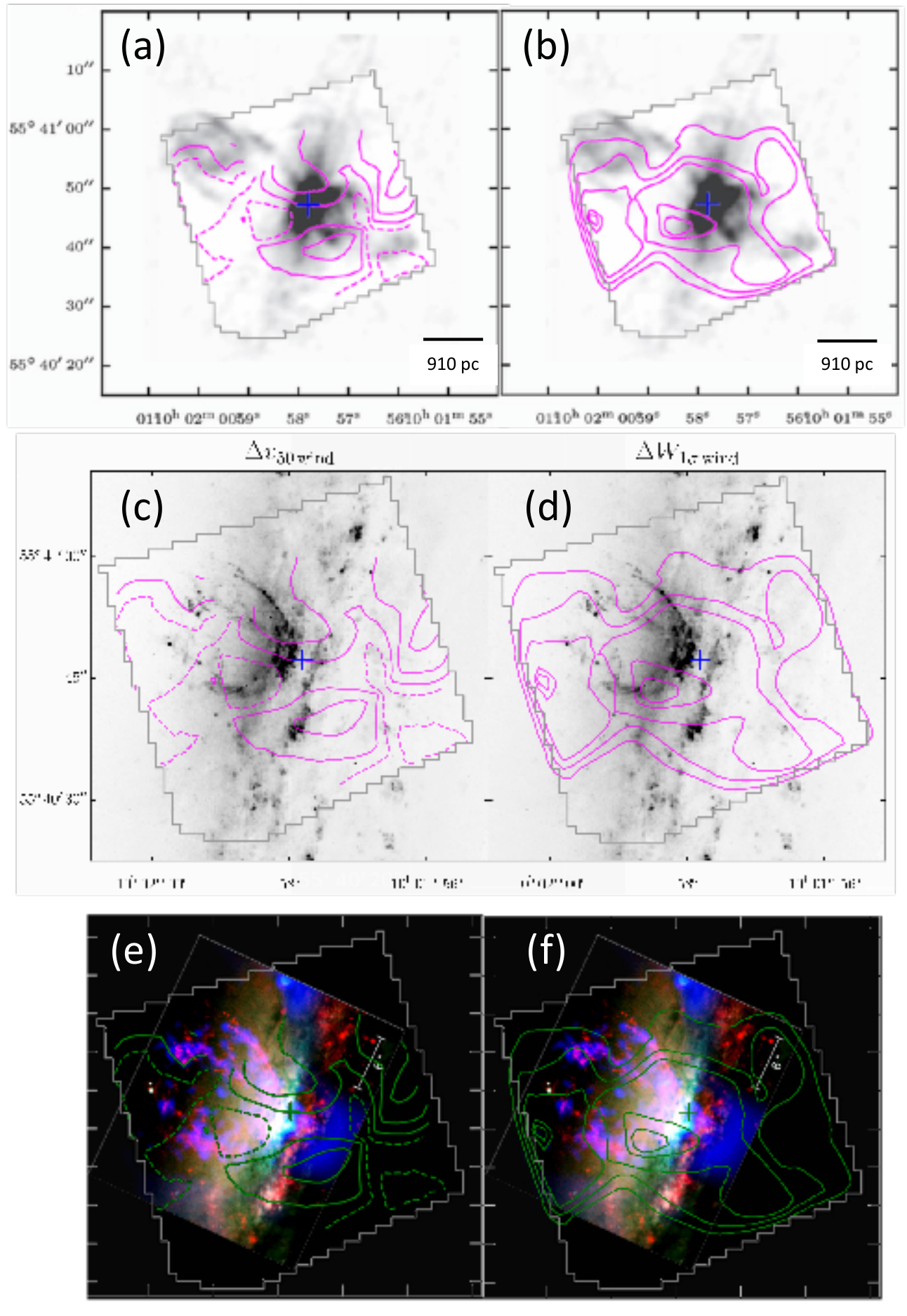}
\caption{PACS contours of the \cii (a) \vFive and (b) \wOne residuals
  in the wind in NGC 3079, overlaid on the 1.4 GHz observations from
  \citet{Sebastian2019}. PACS contours of the \cii (c) \vFive and (d)
  \wOne residuals in the wind in NGC 3079 overlaid on the \ha $+$
        [N~II] image from \citet{Cecil2001}.  PACS contours of the
        \cii (e) \vFive and (f) \wOne residuals in the wind in NGC
        3079 overlaid on the \emph{HST} image from above (red and
        green) and the \emph{Chandra} data (blue) from
        \citet{Cecil2002}. North is up and East to the left.}
\label{fig:ngc3079Wind}
\end{figure*}

Direct comparisons of our results with those at other wavelengths may
help clarify the origin of this excess line broadening.  In the top row
of Figure \ref{fig:ngc3079Wind}, the contours of the \cii
\vFiveDelt\ and \wOneDelt\ residuals are overlaid on the 1.4 GHz image
from \citet{Sebastian2019}, where the double-lobed radio morphology
of the wind is outlined by bright filaments. At the base of the W
lobe, where the radio filaments are brightest, \vFiveDelt\ appears to
trace the lobe's edges. The material along the southern edge is
receding with \vFiveDelt\ $\sim$ $+$35 \kms\ while it is approaching
along the northern edge with \vFiveDelt\ $\sim$ $-$15 \kms.

The middle row of Figure \ref{fig:ngc3079Wind} shows the same contours
as above, but overlaid on the continuum subtracted \ha $+$ [N~II]
image from \citet{Cecil2001}, where only the eastern bubble is visible
out to $\sim$ 1 kpc above the disk \citep[the western bubble lies
  behind the disk of galaxy and is therefore extinguished although not
  completely absent;][]{Veilleux1994}. While there is a lack of radio
emission along the southern edge of the eastern bubble, the strong
optical line emission in this region coincides with approaching \cii
line emitting material with \vFiveDelt\ $\sim$ $-$35 \kms and where
\wOneDelt\ are the largest ($\sim100$ \kms). \citet{Cecil2001} have
shown that the base of this filament aligns with the axis of the jet
observed at 8 GHz \citep{Trotter1998}.
The larger dispersion residuals observed here mimic those seen in the
optical lines \citep{Veilleux1994,Cecil2001} and likely have the same
origin: the jet is interacting with the ISM of the galaxy disk,
depositing its kinetic energy to drive the bubble
\citep{Middelberg2007}. Note however that \wOne\ of H$\alpha$ reaches
up to $\sim$1000 \kms, considerably larger than the [C~II] widths. The
\cii widths are similar to the values derived by \citet{Hawarden1995}
from H$_2$ ro-vibrational transitions and mapped by
\citet{Israel1998}.


Beyond $\sim$ 1 kpc, the southern optical line filament in the eastern
bubble appears to break out into smaller clouds at the location where
the radio emission begins. The values of [C~II] \vFiveDelt\ along this
filament are $\sim$ $-$20 \kms. In contrast, the base of the northern
filament in the eastern bubble is seen in both optical and radio, and
the [C~II]-emitting is receding at velocities \vFiveDelt\ $\sim$ 45
\kms.

The bottom row of Figure \ref{fig:ngc3079Wind} shows the \cii
\vFiveDelt\ and \wOneDelt\ contours overlaid on the \emph{HST} image
from above (red and green) and the \emph{Chandra} data from
\citet{Cecil2002}. Notice that on the East side, the (soft) X-ray
filaments spatially correlate with the optical filaments, while in the
West, the (soft) X-ray emission fills the radio lobe region, but lacks
any of the filamentary structures seen at 1.4 GHz. The
\wOneDelt\ contours are bisymmetric like the X-ray and radio
structures, providing additional support for an association with the
nuclear outflow.

\subsubsection{Overall Influence of the Nuclear Outflow}
\label{sec:otherWind}

The spatial coincidence noted in Sec.\ \ref{sec:multiwavelength}
between the regions with anomalously high \cii line widths (\wOneDelt)
and the well-known visible, radio, and X-ray features associated with
the nuclear outflow in this galaxy points to a physical connection
between the cool gas phase traced by \cii and the warm/hot ionized and
relativisitc plasma components taking part in the outflow. The broad
line profiles of OH 119 and the lack of obvious OH velocity gradient
reported in Sec.\ \ref{sec:results_imaging_spectroscopy_oh} suggest
that the influence of the nuclear outflow also extends to the
molecular gas phase. However, as discussed in Section
\ref{sec:results_imaging_spectroscopy_oh} (Fig.\ \ref{fig:ngc3079oh}),
OH 119 is detected in absorption in only 12-14 spaxels (each 9\farcs4
$\times$ 9\farcs4), or only about 9-10 spatial resolution elements
(10\arcsec\ FWHM), centered on the inner $\sim$ 30\arcsec\ $\times$
30\arcsec\ or $\sim$ 3 $\times$ 3 kpc. The far-infrared continuum
quickly drops beyond this region, making the detection of OH
absorption against this continuum emission more difficult and
revealing possible OH emission. These limitations prevent us from
making a strong statement on the origin of the anomalous OH
kinematics. While we favor an outflow origin for the broad OH profiles
given the similarities with the \cii profiles, we cannot formally rule
out the possibility that some of the OH line broadening is due to bar
streaming or associated with fast rotational velocities in a massive
core, as traced by CO (1-0) \citep{Koda2002}.


Overall, our analysis of the PACS data indicates that the cool gas
traced by \cii in the nuclear region of this galaxy is influenced by
the ionized + relativistic outflow. This component of the ISM has been
found in some gas-rich systems to dominate the mass budget of galactic
outflows and sometimes even their energetics \citep[e.g.,][and
  references therein]{Veilleux2020}. This does not seem to be the case
in NGC~3079 although the limited spatial and spectral resolutions of
the PACS data prevent us from carrying out a detailed dynamical
analysis of the [C~II] outflow.  In particular, the data cannot be
used to reliably decompose the [C~II] flux into an outflow component
and a disk component.
The velocities of the cool gas traced by \cii are modest
compared with the escape velocity. In most circumstances \citep[e.g.,
][]{Veilleux2020}, the escape velocity in a disk galaxy can be
approximated as $v_{\rm esc} \simeq (2 - 3) v_{\rm rot}$, where
$v_{\rm rot}$ is maximum rotation speed. Kinematic modeling of the
large-scale H$\alpha$ velocity field in NGC~3079 \citep{Veilleux1999}
gives $v_{\rm rot} = 250$ \kms, so $v_{\rm esc} = 500 - 750$ \kms. The
bulk of the outflowing cool gas will thus take part in a galactic
fountain that reaches the inner CGM rather than be part of a genuine
wind that escapes the host galaxy altogether. This is also likely to
be the case for the neutral-gas phase traced by the blueshifted H~I 21
cm absorption \citep{Shafi2015}.

While direct mechanical feedback by the cool outflow on the ISM may be
limited to the inner portion of the galaxy CGM, the elevated
100-to-160 $\mu$m flux ratios of Figure
\ref{fig:ngc3079_pacs_100_160_ratio}, coincident with the brightest
X-ray emission, suggest that the influence of radiative feedback may
extend much further. The elevated PAH 7.7/11.3 $\mu$m ratios found by
\citet{Yamagishi2010} add support to this idea. The outflowing cool
gas may undergo a phase transition, becoming ionized as it travels to
the halo, and may contribute to the warm and hot ionized outflows that
are detected on larger scales.



\section{Conclusions}
\label{sec:conclusions}

We have presented the results from a photometric analysis of deep {\em
  Herschel} PACS and SPIRE far-infrared images of the nearby edge-on
disk galaxy NGC~3079, host to a well-known AGN-driven outflow. We also
carried out a kinematic analysis of the \cii line emission and OH 119
line absorption within the central $\sim$ 6 kpc region of this galaxy
derived from PACS data cubes. The main results from these analyses are
the following:

\begin{itemize}
\item The PACS images at 100 $\mu$m (PSF FWHM $\simeq$ 6\farcs8) and
  160 $\mu$m (PSF FWHM $\simeq$ 11\farcs4) reveal a distinct X-shape
  structure that extends over 25 $\times$ 25 kpc$^2$ centered on the
  nucleus. The filaments connect back to the disk of NGC~3079, about 5
  kpc on each side of the nucleus. One of the filaments is detected
  out to $\sim$ 25 kpc from the nucleus and $\sim$ 15 kpc from the
  galaxy disk mid-plane.

\item Dust temperatures of 16-18 K and a total dust mass of (4 $-$ 16)
  $\times$ 10$^6$ $M_\odot$ are derived from fits to the far-infrared
  spectral energy distributions from the individual filaments making
  up the X-shape structure. A gas mass of (8 $-$ 30) $\times$ 10$^8$
  $M_\odot$ is implied if the global gas-to-dust ratio of 200 measured
  from the total gas and dust masses in this galaxy also applies to
  the X-shape structure.

\item Comparisons with published data at other wavelengths suggest
  that the far-infrared X-shape structure is physically associated
  with the large-scale galactic wind detected in H$\alpha$, X-rays,
  and far-ultraviolet. In this picture, dusty material originally in
  the galactic disk of NGC~3079 is lifted above the disk and entrained
  in the galactic wind. The energy needed to lift this material,
  $\sim$ 10$^{56}$ ergs, may have been supplied by the central AGN
  provided that it has maintained the same level of activity for the
  past $\sim$ 10$^7$ yrs. A large mass-loading factor ($\ga$ 100)
    is implied in this scenario.

\item Additional support for this scenario is found in the central 10
  $\times$ 10 kpc region.  Elevated 100-to-160 $\mu$m flux ratios,
  indicative of higher dust temperatures, are observed within a
  biconical region centered on the nucleus, coincident with the
  brightest soft X-ray emission. A similar coincidence between warm
  dust and X-ray emission has been reported by our group in NGC~4631
  (Paper I) and NGC~891 (Paper III). In NGC~3079, dust inside the
  bicone may be heated by UV radiation from the AGN, X-ray emission
  from the hot plasma in the halo, or shocks associated with the
  large-scale outflow into the halo.

\item To survive, the dust must be shielded from fast shocks. A
  thermal sputtering timescale of $\sim$10$^7$ yrs is derived if the
  hot X-ray emitting material in the X-shape filaments is in contact
  with the dust. This timescale is similar to the duration of the AGN
  jet cycle needed to produce the far-infrared X-shape structure.

\item A careful analysis with \barolo\ of the [C~II] kinematics in the
  inner 6 $\times$ 6 kpc$^2$ derived from the PACS data cube reveals
  line broadening along the minor axis of this galaxy that is well in
  excess of the values expected from beam smearing of the disk
  rotational motion.  The region of excess line broadening coincides
  loosely with the bisymmetric superbubbles and filaments seen in
  H$\alpha$, soft X-rays, and at radio wavelengths. This is
  interpreted as a sign that the nuclear warm/hot ionized +
  relativistic outflows in this object is stirring the cool gas
  traced by [C~II]. The disturbed OH kinematics in this same region
  suggest that the molecular gas is also influenced by the nuclear
  outflow, although the presence of a bar and compact nuclear core
  makes this conclusion less certain than in the case of the cool gas.
\end{itemize}

The {\em Herschel} far-infrared data trace the complex interplay of
the cool gas/dust with the warm/hot ionized + relativistic
phases of the large-scale outflow detected at other wavelengths. The
cool gas entrained in the inner hot/relativistic outflow has modest
velocities relative to the escape velocity of the host, so it will
eventually be deposited in the halo of the galaxy and contribute to
building up the gas, dust, and metal contents of the inner CGM.  The
exact fate of this material is not constrained by our data. It may
experience a phase transition from neutral to ionized due to AGN
photoionization and/or shocks as it propagates outward into the
halo. However, without a sustained source of energy from the AGN or
nuclear starburst, the material in the bicone will cool down and rain
back onto the galaxy disk, closing the loop of a large-scale galactic
fountain. Deeper and higher resolution infrared data of this and other
nearby galaxies with the upcoming {\em James Webb Space Telescope} and
planned {\em Origins Space Telescope} will provide unique insights
into this disk-halo gas circulation pattern which is believed to
regulate the evolution of disk galaxies such as our own.

\section*{Acknowledgements}

We thank the anonymous referee for suggestions which improved
  this paper. We are grateful to H\'el\`ene Roussel for her help with
the use of {\em Scanamorphos}. This work was supported in part by the
National Science Foundation (NSF) under AST-1009583 (SV) and
ASTR-1817125 (CLM), JPL Awards 1276783 and 1434779, and NASA grants
NHSC/JPL RSA 1427277, 1454738 (SV and MM), and ADAP NNX16AF24G (SV and
MS). This work has made use of NASA's Astrophysics Data System
Abstract Service and the NASA/IPAC Extragalactic Database (NED), which
is operated by the Jet Propulsion Laboratory, California Institute of
Technology, under contract with the National Aeronautics and Space
Administration. PACS has been developed by a consortium of institutes
led by MPE (Germany) and including UVIE (Austria); KU Leuven, CSL,
IMEC (Belgium); CEA, LAM (France); MPIA (Germany);
INAF-IFSI/OAA/OAP/OAT, LENS, SISSA (Italy); IAC (Spain). This
development has been supported by the funding agencies BMVIT
(Austria), ESA-PRODEX (Belgium), CEA/CNES (France), DLR (Germany),
ASI/INAF (Italy), and CICYT/MCYT (Spain). SPIRE has been developed by
a consortium of institutes led by Cardiff University (UK) and
including Univ.\ Lethbridge (Canada); NAOC (China); CEA, LAM (France);
IFSI, Univ.\ Padua (Italy); IAC (Spain); Stockholm Observatory
(Sweden); Imperial College London, RAL, UCL-MSSL, UKATC, Univ.\ Sussex
(UK); and Caltech, JPL, NHSC, Univ.\ Colorado (USA). This development
has been supported by national funding agencies: CSA (Canada); NAOC
(China); CEA, CNES, CNRS (France); ASI (Italy); MCINN (Spain); SNSB
(Sweden); STFC, UKSA (UK); and NASA (USA). HIPE is a joint development
by the {\em Herschel} Science Ground Segment Consortium, consisting of
ESA, the NASA {\em Herschel} Science Center, and the HIFI, PACS and
SPIRE consortia.

\section*{Data Availability}

The data underlying this article will be shared on reasonable request
to the corresponding author.




\bibliographystyle{mnras}
\bibliography{references} 

\bsp	
\label{lastpage}
\end{document}